\documentclass[journal]{IEEEtran}

\ifCLASSINFOpdf
\else
\fi
\usepackage{slashbox,pict2e}
\usepackage{xcolor}
\usepackage{float}
\usepackage{stfloats}
\usepackage{graphicx}
\usepackage{lineno,hyperref}
\modulolinenumbers[5]
\usepackage{amsmath}
\usepackage{epstopdf}
\usepackage{graphicx,cite,multirow,rotating}
\usepackage{setspace}
\usepackage{pgfplots} 
\usepackage{tikz}
\usepackage{enumitem}

\usetikzlibrary{shapes,arrows}
\usepackage{varwidth}
\usetikzlibrary{shapes.geometric, arrows}
\usetikzlibrary{positioning,arrows}

\usepackage{amssymb}
\usepackage{amsfonts}
\usepackage{yfonts}
\usepackage{psfrag}
\usepackage{pifont}
\usepackage{amsfonts}
\usepackage{bbm}
\usepackage{dsfont}

\usepackage{color}
\usepackage{amsmath,stackrel,dsfont}
\usepackage{amssymb,hyperref,stmaryrd}
\usepackage{algorithm}
\usepackage{algpseudocode}
\usepackage{pifont}
\modulolinenumbers[5]
\usepackage{amsmath}
\usepackage{epstopdf}
\usepackage{graphicx,cite,multirow,rotating}
\usepackage{setspace}
\usepackage{pgfplots} 
\usepackage{tikz}

\usetikzlibrary{shapes,arrows}
\usepackage{varwidth}
\usetikzlibrary{shapes.geometric, arrows}
\usetikzlibrary{positioning,arrows}
\usepackage{amssymb}
\usepackage{amsfonts}
\usepackage{yfonts}
\usepackage{psfrag}
\usepackage{pifont}
\usepackage{amsfonts}
\usepackage{bbm}
\usepackage{dsfont}

\usepackage{color}
\usepackage{amsmath,stackrel,dsfont}
\usepackage{amssymb,hyperref,stmaryrd}
\usepackage{algorithm}
\usepackage{algpseudocode}
\usepackage{pifont}

\ifCLASSOPTIONcompsoc
\usepackage[caption=false, font=normalsize, labelfont=sf, textfont=sf]{subfig}
\else
\usepackage[caption=false, font=footnotesize]{subfig}
\fi

\begin{document}
%
\title{Fast and Robust LRSD-based SAR/ISAR Imaging and Decomposition}
%
%
%

\author{Hamid~Reza~Hashempour, Majid~Moradikia, Hamed~Bastami, Ahmed~Abdelhadi, Mojtaba Soltanalian, \textit{Senior Member, IEEE}
\thanks{Hamid Reza Hashempour is with the School of Electrical and Computer Engineering, Shiraz University, Shiraz,
Iran (e-mail: hrhashempour@shirazu.ac.ir.)

Majid Moradikia and Ahmed Abdelhadi are with the Department of Engineering Technology, University of Houston, Houston, USA (e-mail: mmoradik@central.uh.edu; aabdelhadi@uh.edu.)

Hamed Bastami is with the Department of Electrical Engineering, Sharif University of Technology, Tehran,
Iran, (e-mail: hamed.bastami@ee.sharif.edu)

Mojtaba Soltanalian is with the Department of Electrical and Computer
Engineering, University of Illinois at Chicago, USA (e-mail: msol@uic.edu).
This work was supported in part by the National Science Foundation Grant ECCS-1809225.
}}

%
%

\markboth{IEEE ,~Vol.~?, No.~?,  ~}%
{Shell \MakeLowercase{\textit{et al.}}: Bare Demo of IEEEtran.cls for IEEE Journals}
%



\maketitle

\begin{abstract}
The earlier works in the context of low-rank-sparse-decomposition (LRSD)-driven stationary SAR imaging have shown significant improvement in the reconstruction-decomposition process. Neither of the proposed frameworks, however, can achieve a satisfactory performance when facing a platform residual phase error (PRPE) arising from the instability of airborne platforms. More importantly, in spite of the significance of real-time processing requirement in the remote sensing applications, these prior works have only focused on enhancing the quality of the formed image, not reducing the computational burden. To address these two concerns, this paper presents a fast and unified joint SAR imaging framework where the dominant sparse objects and low-rank features of the image background are decomposed and enhanced through a robust LRSD. In particular, our unified algorithm circumvents the tedious task of computing the inverse of large matrices for image formation and takes advantages of the recent advances in constrained quadratic programming to handle the unimodular constraint imposed due to the PRPE.  Furthermore, we extend our approach to ISAR autofocusing and imaging. Specifically, due to the intrinsic sparsity of ISAR images, the LRSD framework is essentially tasked with the recovery of an sparse image. Several experiments based on synthetic and real data  are presented to validate the superiority of the proposed method in terms of imaging quality and computational cost as compared to the state-of-the-art methods. 
\end{abstract}


\IEEEpeerreviewmaketitle

\section{Introduction}\label{intro}
\IEEEPARstart{S}{ynthetic} Aperture Radar (SAR) is an active remote sensing system which can capture high-resolution images of the terrain \cite{Cumming}. A key challenge in remote sensing applications is to facilitate SAR imaging in real-time processing\cite{Moreira,Cantalloube,Zhou-Real-Time,Hashem_FADMM,Asadipooya}. Such a real-time processing in stripmap mode is defined by the time duration before the next synthetic aperture data is received, which is about several seconds \cite{Cumming,Moreira}.

The SAR processor works to reconstruct the target image from the backscattered data. The conventional SAR image formation schemes suffer from several practical limitations such as available system bandwidth, speckle, and side-lobe artifacts \cite{Hashem_FADMM,Asadipooya,Samadi,Dong,Bu,Onhon,Chen-SAR,Soganli,moradikia}. To address such shortcomings, \cite{Samadi} incorporated the sparsity assumption into the SAR imaging problem as a prior. 
In \cite{Dong} and \cite{Bu}, the authors exploit the compressed sensing (CS) theory for SAR imaging with the undersampled data.
In practice, the platform fluctuations perturb the received signal, in such a way that even after motion compensation, the platform residual phase errors (PRPE) remains in the received signal at various cross-range positions \cite{Onhon}. The authors of \cite{Onhon} have aimed at addressing this issue by proposing an autofocus SAR imaging method through which the sparse nature of the scene is enhanced while the PRPE is corrected for simultaneously. However, the above methods have no mechanism to deal with non-smooth/non-sparse patterns involved in the scene which can be represented as low-rank structures \cite{Soganli,moradikia}. 

To concurrently enhance and separate both the sparse and low-rank features of the scene, the low-rank and sparse decomposition (LRSD) model has been leveraged for SAR imagery in \cite{Soganli,moradikia,Video}. Before applying an LRSD framework, the signal sparsity is highlighted through an appropriate dictionary while at the same time the low-rankness is captured by constructing the so-called Casorati matrix from overlapping SAR image patches \cite{moradikia}. To have a successful decomposition via LRSD, there must only be slight differences among various columns of the received data at different cross-range positions \cite{Video}. More explicitly, in the presence of PRPE, the existing LRSD-SAR schemes can no longer work properly. In \cite{Video}, we have integrated the PRPE correction (PRPEC) step with the LRSD technique to facilitate decomposing the sparse signatures of moving objects along the temporal consecutive video-frames from the static background. Although including the PRPEC in static LRSD-based SAR imaging has not been studied yet, motivated by the discussion in  \cite{Video}, it can be intuitively inferred that the LRSD-based approach of \cite{moradikia} cannot achieve a satisfactory decomposition result in the presence of realistic PRPE assumption. 

Note that, neither of the SAR imaging approaches mentioned above have focused on reducing the computational burden, but rather enhancing the quality of output image. For instance, the algorithms proposed in \cite{Onhon,Chen-SAR,Soganli,moradikia,Video} rely on specific gradient-based approaches, whose application would be problematic in large-scale data setting, due to the tedious task of calculating large matrix inversions. Such difficulties would be exacerbated in case the data is stored on various computers \cite{Asadipooya}. Additionally, the approaches in \cite{Hashem_FADMM,Asadipooya,Samadi,Dong,Bu,Onhon,Chen-SAR,Soganli,moradikia,Video}, involve the stacking of two-dimensional matrices into large vectors, and large dictionary which is not efficient in terms of required storage and computational loads. Furthermore, the conventional alternating direction method of multiplier (ADMM) steps used in \cite{moradikia} and \cite{Video} introduce some additional auxiliary steps which slow down the emerging algorithm. 

High resolution inverse synthetic aperture radar (ISAR) autofocusing and imaging have been widely studied in recent years \cite{Chen,Zhang,Qiu,Liu,Zhao,HashempourDSP,S_Zhang,Hashempour_sparsity_driv,admm-sbl}. ISAR images are known to be inherently sparse \cite{HashempourDSP}. Therefore, sparsity-driven methods have emerged for ISAR image reconstruction. However, similar to the SAR problem, conventional sparse representation-based methods demand large memory and heavy computation loads, which is a significant challenge for real-time processing \cite{Qiu}. Thus, the sparse matrix models are exploited in \cite{Qiu,Hashempour_sparsity_driv} for faster ISAR reconstruction. Specifically, a fast ADMM based algorithm was used in \cite{Hashempour_sparsity_driv} to reduce the imaging complexity. The authors of \cite{Zhao} introduce a sparse Bayesian learning (SBL) method for ISAR autofocusing and imaging. On the other hand, matrix inversion computations are required in this approach, which renders it infeasible for real time scenarios. Thus, \cite{admm-sbl} exploits ADMM steps and introduces an ADMM-based SBL (ADMM-SBL) algorithm to do away with the matrix inversion and reduce the computational cost. However, the main drawback of ADMM-SBL is the computation of the first and the second derivative of entropy which is still time consuming.

Interestingly, due to the unimodular nature of PRPE, the PRPEC creates an NP-hard problem known as a unimodular quadratic program (UQP), where a quadratic form is optimized over the unimodular vector set. Given the recent advances in solving such kinds of problem \cite{Soltanalian1,Soltanalian2,Soltanalian3}, this paper proposes a fast unified joint SAR imaging framework where sparse dominant objects and low-rank features of the background are decomposed and enhanced through a robust LRSD. In particular, our unified proposed algorithm can deal with the challenging matrix inversion tasks and speeds up the process through recent advances in approximating UQP solutions, while keeping the matrix forms of SAR kernel rather than stacking them into inefficient vector forms. Another appealing advantage of the proposed iterative approach is that each iteration is given through a closed-form formula that facilitates simple and immediate implementation. This idea is also extended to ISAR autofocusing and imaging. The main difference in this case is that the ISAR image is not low-rank. Consequently, the task of the LRSD framework is reduced to sparse recovery.

The rest of this paper is organized as follows.  A brief review of the previous related works is described in Section \ref{Related}.
Section \ref{pre}
introduces the observation model and the conventional ADMM-based approach. Section \ref{propsed} presents the proposed fast SAR/ISAR imaging method. Then, several experiments based on synthetic and real data are performed to validate the efficiency and superiority of the proposed approach in Section \ref{experimet}. Finally, Section \ref{conc} concludes the paper.

\textit{Notation}: We use bold lowercase letters for vectors and bold uppercase letters for matrices. $(\cdot)^T$ and $(\cdot)^H$  denote the transpose operator and the conjugate transpose operator, respectively. $\odot$ and $\otimes$ denote the Hadamard product and Kronecker product, respectively. $\mathbf{1}$ is the all-ones vector/matrix.
 $\mathrm{Vec(\cdot)}$ stands for stacking an $M \times N$ matrix into an $MN \times 1$ vector column by column, and $\mathrm{Mat(\cdot)}$ creates a matrix from vector and actually is the inverse operator of $\mathrm{Vec(\cdot)}$. $\mathrm{diag}\{\cdot\}$ denotes constructing a diagonal matrix from a vector.
For a given matrix $\mathbf{A}$, $\mathbf{A}_{i,j}$ denotes its  $(i,j)$-th element.

\section{Related Works}\label{Related}
Before presenting the signal model and our approach to dealing with LRSD-based radar imaging,
we briefly review some related works in radar imaging as well as the previously developed fast LRSD algorithms.

\subsection{Applications of LRSD in Radar Imaging}\label{Application of LRSD}
Recently, the LRSD approach also known as robust principal component analysis (RPCA) has been widely adopted in the SAR community for various applications. In the previous section, the application of LRSD in SAR imaging and decomposition was briefly introduced and some recent works \cite{Soganli,moradikia,Video} were taken into consideration. Besides, RPCA has been applied in the other SAR subjects, such as ground moving target indication (GMTI) SAR  (SAR-GMTI) \cite{GMTI1,GMTI2,GMTI3,GMTI4,GMTI5,GMTI6}. In SAR-GMTI, the clutter is static and has approximately no changes between successive SAR images. Therefore, it can be considered as the low-rank part. On the other hand, due to the Doppler effect, the moving target echo changes in sequential images and is the sparse part to be recovered. Therefore, the SAR-GMTI approach can be interpreted as LRSD in which the goal is to extract the moving target, i.e. the sparse part, from the SAR images. It is worth noting that multiple images of the scene can be provided simultaneously by a multichannel SAR system or in successive time durations by a one-channel system. In \cite{GMTI1}, an approach for extracting
moving targets in a multichannel wide-area surveillance radar system is introduced. In the proposed algorithm, after related preprocessing, the radar echoes are combined in a matrix as the superposition of three matrices, namely, a low-rank
matrix of ground clutter, a sparse matrix of moving targets,
and an entry-wise matrix of noise component. Then, the relaxed principal component pursuit (PCP) is used to separate the ground clutter (low-rank matrix) and the moving target (sparse matrix). The ref. \cite{GMTI2} is another work in multichannel SAR-GMTI that provides an efficient along-track interferometry (ATI) go decomposition (GoDec) approach for GMTI under a strong clutter background. GoDec algorithm separates the low-rank and sparse matrices from
the original matrix by tackling the RPCA basic model.
Even though the GoDec algorithm runs faster than the
the augmented Lagrange multiplier
(ALM) methods, it still needs a long time to iterate for better performance \cite{GMTI2}. Authors in \cite{GMTI3}  use LRSD for maritime surveillance. Specifically, their goal is to separate the sparse objects of interest, i.e. maritime targets, from the stationary low-rank background by solving the RPCA problem via convex programming. 
Forward-looking scanning radar is another application that can utilize LRSD for multiple-target extraction from the real-beam image \cite{GMTI4}. 

In LRSD, the problem of recovering the low-rank part is usually transformed into a minimization of the nuclear-norm, i.e.  the sum of singular values
of the original matrix. The singular value thresholding (SVT), is a proximal
mapping corresponding to the nuclear-norm. However, it is computationally expensive due to the calculation of the singular
value decomposition \cite{GMTI5}. Therefore, \cite{GMTI5} exploits a rank-1 framework instead of SVT and provides an updated minimization problem for LRSD. However, the main drawback of this method is the non-convexity of the rank-1 constraint. As a result, the convergence of the algorithm is not guaranteed. In \cite{GMTI6}, a joint multi-channel sparsity approach to RPCA is intoduced for SAR-GMTI by enhancing the performance of clutter
suppression and improving the sparse signature
of moving targets.

RPCA is also used in SAR peckle reduction problem for the circular SAR system \cite{Speckle},
in which the motion of aircraft platform causes successive 
angular variations so that multiple SAR images can be reached
with a high interrelationship \cite{Speckle}. The authors in \cite{Speckle} develop a robust  $l_p$-regularized scheme instead of RPCA to take into account the low-rank property of targets with a
flexible choice of $1 \leq p \leq 2$ for sparsity promotion to obtain a cleaner target structure while preseving the edge features of the target.
The ALM framework  is then applied to solve the joint
optimization problem with efficient computation of each
ALM subproblem.

\subsection{Current Fast Algorithms for LRSD}\label{Fast Algorithms of LRSD}

In recent years, several fast algorithms for solving the LRSD problem have been introduced \cite{Fast-LRSD1,Fast-LRSD2,Fast-LRSD3,Fast-LRSD4,Fast-LRSD5,Fast-LRSD6,Fast-LRSD7} that will be briefly described in the following.

In \cite{Fast-LRSD1}, the authors surrogate singular value decompositions (SVD) of the data matrix, which has a high computational cost, with the so-called $l_1$-filtering, to
exactly solving PCP. The $l_1$-filtering approach not only is highly parallelizable but also can solve a nuclear norm minimization problem exactly in linear time, which facilitates the application of RPCA to extremely large-scale problems. The ref.
 \cite{Fast-LRSD2} introduces a fast incremental RPCA (FRPCA) approach, in which the low-rank matrices of the incrementally-observed data are estimated
using a convex optimization model.
This model exploits information acquired from the pre-estimated low-rank matrices
of the original observations and can be used in specific applications with real-time characteristics such as background
subtraction in video streams.

Yang \cite{Fast-LRSD3} utilized an smoothing technique to smooth the non-smooth terms in the objective function, and presented a fast alternating direction method for solving RPCA. 
In order to avoid the high computation cost to compute a SVD in SVT,
Oh \textit{et al.} \cite{Fast-LRSD4} surrogate SVT with the fast
randomized SVT (FRSVT), where the direct computation of SVD is no longer needed. Moreover, FRSVT can reduce the computational time of low-rank applications without losing accuracy and hurting the convergence behavior. 
A Normalized Coherence Pursuit (NCP) method was introduced by \cite{Fast-LRSD5} to solve RPCA and was shown (theoretically and numerically) to be
robust to different types of outliers. Zhang \textit{et al.} \cite{Fast-LRSD6}, extended the proximal gradient methods (PGMs) to the proximal
Jacobian iteration methods (PJIMs) for handling with a family of nonconvex composite optimization problems with two splitting variables. The authors showed that PJIMs not only keep fast convergence speed but also have a high precision. Finally, \cite{Fast-LRSD7}  presented a new convergent parallel splitting ALM (PSALM) for two block separable convex programming, which is the regularizing ALM’s minimization subproblem by some simple proximal terms. This new PSALM can be used to solve video background extraction problems efficiently.

To summarize what we have discussed, it can be understood that the recent fast LRSD methods focus mainly in exchanging the conventional algorithms for solving RPCA with new fast approaches. Specifically, SVT, ALM or PGM methods are replaced with new faster counterparts. However, none of these works target the RPCA problem in SAR applications. By applying the SAR kernel in the LRSD problem, as we will discuss in the next sections, we have to compute matrix inversions that are time consuming. On the other hand, the SAR signal model in the LRSD problem is usually written in the vector form which is not efficient. Finally, dealing with PRPE in the LRSD-based SAR imaging and decomposition is a challenging task. Therefore, our goal is to tackle these problems in a computationally efficient way as discussed in detail in the following sections.

\section{Preliminaries}\label{pre}
In this section, after presenting the SAR imaging model, the LRSD principle, as well as the LRSD-based SAR image formation based on the work in \cite{moradikia}, are reviewed. However, we consider a slight extension as compared to \cite{moradikia}, where the practical impact of uncertainties due to platform motion is also taken into account and compensated for.

\subsection{Observation Model}
Using a spotlight mode, a chirp pulse is transmitted at equal angular intervals to illuminate a single spot of a terrain. After applying the far-field approximation, the preprocessing steps of de-chirping and low-pass filtering, as well as two dimensional under-sampling \cite{Hashempour_sparsity_driv}, the discrete form of the back-scattered signal in the presence of realistic phase error due to PRPE, is formulated as follows \cite{Video,Cetin1}:

\begin{equation}\label{EQ2}
\mathbf{r}= \mathbf{P\Theta Hf}+\mathbf{n}
\end{equation}
in which $\mathbf{r}\in {\mathbb{C}}^M$ is comprised of the undersampled phase histories corresponding to azimuth angles ${\{\theta }_i\}^{N_a}_{i=1}$,  $\mathbf{H}\in {\mathbb{C}}^{N\times I}$ is the discretized approximation of the observation kernel with $N=N_a \times N_r$  where $N_a$ and $N_r$, respectively, stand for the number of full-sampled data at  azimuth and range directions, and $I$ is the number of pixels in the scene. Additionally,
\begin{align}\label{Eq-phi}
\mathbf{P}&\triangleq  \mathrm{diag} \{  \mathbf{p}  \} 
\nonumber \\ & =
 \mathbf{I}_{M_r} \otimes
\boldsymbol{\Phi}
\end{align}
 with $\boldsymbol{\Phi}=\mathrm{diag}\left\{e^{j{\phi }_1},e^{j{\phi }_2},\dots ,e^{j{\phi }_{M_a}}\right\}\in {\mathbb{C}}^{M_a \times M_a}$
 containing the exponentiated PRPE values, and $\mathbf{I}_{M_r}$ is an $M_r \times M_r$ identity matrix.  The vector $\mathbf{f}\in{\mathbb{C}}^{I\times 1}$ represents the vectorized form of the unknown reflectivity image. Finally, $\boldsymbol{\Theta }\in{\mathbb{C}}^{M\times N}$(with $M=M_a M_r\ll N$ where $M_r$ and $M_a$ respectively stand for the number of undersampled data at range and azimuth directions) and $\mathbf{n} \in{\mathbb{C}}^M$ denote the sensing matrix and the undersampled Gaussian noise, respectively. Note that, $\mathbf{H}$ can be viewed as the 2-D Fourier transform of the reflectivity field whose support region constitutes a sectorial part of an annulus \cite{Video,moradikia}.

\subsection{LRSD-Based SAR Image Formation}
The LRSD approach decomposes the hypothetical matrix $\mathbf{F}$ into its low rank $\mathbf{L}$ and sparse $\mathbf{S}$ components by solving the following convex optimization problem \cite{Candes,Vaswani1,Vaswani2,Sobral}:
\begin{equation}\label{EQ3}
{\mathop{\mathrm{min}}_{\mathbf{L},\mathbf{S}}  \ {\left\|\mathbf{L}\right\|}_*+\lambda {\left\|\mathbf{S}\right\|}_1\ }\ \ \ \ \ \  \ \ \mathrm{s.t.}\ \ \ \ \mathbf{F}=\mathbf{L}+\mathbf{S}
\end{equation}
where, ${\left\|\cdot\right\|}_*$ represents the nuclear norm and ${\left\|\cdot\right\|}_1$ denotes the $l_1$-norm of the matrix argument. It has been readily shown in \cite{Candes} the decomposition via LRSD is unique and the problem is well-posed if the matrix $\mathbf{F}$ is truly composed of low rank ($\mathbf{L}$) and sparse ($\mathbf{S}$) components.
In a complete SAR image, while sparsity is caused by the dominant point scatterers, the low-rank presumption is not entirely realistic. In order to make the model more accurate, and to highlight the inherent low-rankness embedded in the scene, a patch-based model can be used \cite{moradikia}. To do so, using the linear operator $\mathcal{G}$ a patch-based matrix $\mathbf{F} \in \mathbb{C}^{n\times K}$ is built from the image $\mathbf{f}$, i.e., $\mathbf{F}=\mathcal{G}\left\{\mathrm{Mat}\left\{\mathbf{f}\right\}\right\}=\mathcal{P}\left\{\mathbf{f}\right\}$, such that the emerging matrix has both sparse and low rank components, i.e. $\mathbf{F}=\mathbf{L}+\mathbf{S}$. The dimension of the obtained matrix $\mathbf{F}$ is determined by the number of sliding steps of the considered sliding window, denoted by $K$, and the number of pixels included within the sliding window, i.e., $n$. In the reconstruction process, we can utilize an inverse operator\footnote{We have omitted the elaborate derivations associated with the operators $\mathcal{G}$ and  ${\mathcal{G}}^{-1}$ here due to space constraints. The interested reader, however, may refer to \cite{moradikia} for details.} ${\mathcal{P}}^{-1}\left\{.\right\}$, such that $\mathbf{f}=\mathrm{Vec}\left\{{\mathcal{G}}^{-1}\left\{\mathbf{F}\right\}\right\}={\mathcal{P}}^{-1}\left\{\mathbf{F}\right\}$. 
It is worth noting that	the parameter $\lambda$ in \eqref{EQ3} is the regularization parameter which plays an important role in the trade-off between the low-rankness and sparsity of $\mathbf{L}$ and $\mathbf{S}$, respectively. Candes et al. \cite{Candes} suggested the theoretically supported value of $\lambda = \max (n,K)^{-1/2}$.

Using the observation model in \eqref{EQ2}, the LRSD-SAR imaging framework can then be formulated via a coordinate decent approach where in the first step, considering known $\mathbf{p}$, the following optimization problem has to be solved
\begin{equation}\label{EQ4}
\left[\widehat{\mathbf{L}},\widehat{\mathbf{S}}\right]={\mathrm{arg} {\mathrm{min} \ \frac{1}{2}{\left\|\mathbf{r}-\mathbb{E}\left\{\mathbf{L}+\mathbf{S}\right\}\right\|}^2_2+{\lambda }_L{\left\|\mathbf{L}\right\|}_*+{{\lambda }_s\left\|\mathbf{S}\right\|}_1\ }\ }
\end{equation}
where $\mathbb{E}\left\{\cdot\right\}{\triangleq }\mathbf{P\Theta H}{\mathcal{P}}^{-1}\left\{\cdot\right\}$, and  ${\lambda }_L$ and ${\lambda }_s$ are the regularization parameters. In the next step, using the matrix parameters $\mathbf{L}$, $\mathbf{S}$ acquired from \eqref{EQ4}, and by reformulating \eqref{EQ2} in the form $\mathbf{r}=\boldsymbol{\mathfrak{T}} \mathbf{p}+\mathbf{n}$ with $\boldsymbol{\mathfrak{T}}\triangleq \mathrm{diag}\left\{\boldsymbol{\mathrm{\Theta }}\mathbf{H}\mathbf{f}\right\}$, an estimate of vector $\mathbf{p}$ can be found through the following minimization problem:
\begin{equation}\label{EQ5}
\widehat{\mathbf{p }}={\mathrm{arg} {\mathop{\mathrm{min}}_{\mathbf{p }} \ {\left\|\mathbf{r}-\boldsymbol{\mathfrak{T}}\mathbf{p}\right\|}^2_2 }}\ \ \ \ \mathrm{s.t.}\ \ \ \ \left|{\left(\mathbf{p }\right)}_i\right|=1,\ \ \ \ \forall i. 
\end{equation} 

In order to solve \eqref{EQ4} an algorithm based on ADMM, was developed in \cite{moradikia}. In particular, auxiliary variables of $(\mathbf{W},\mathbf{Q})$ were introduced to mirror $(\mathbf{L},\mathbf{S})$, through which the associated augmented Lagrangian function is minimized with respect to each of $\mathbf{W},\mathbf{Q},\mathbf{L}$, $\mathbf{S}$ separately. Namely, the estimates of $\mathbf{L}$ and $\mathbf{S}$ at the $\left(k+1\right)$th iteration are obtained as follows \cite{moradikia}:

\begin{equation}\label{EQ6}
{\widehat{\mathbf{W}}}^{k+1}={\mathrm{Prox}}_{({{\lambda }_L}/{{\delta }_{1}}){\left\|.\right\|}_*}\left\{{\mathbf{F}}^k -{\mathbf{S}}^k+{\delta }^{-1}_1{\mathbf{Z}}^k_1\right\},
\end{equation}
\begin{equation}\label{EQ7}
{\widehat{\mathbf{Q}}}^{k+1}={\mathrm{Prox}}_{({{\lambda }_s}/{{\delta }_{2}}){\left\|.\right\|}_1}\left\{{\mathbf{F}}^k-{\mathbf{L}}^k+{\delta }^{-1}_2{\mathbf{Z}}^k_2\right\},
\end{equation}
\begin{align}\label{EQ8}
{\widehat{\mathbf{L}}}^{k+1}=&\underbrace{{\left({{\mathbb{E}}^H}^k{\mathbb{E}}^k+{\delta }_{1}\mathbf{I}\right)}^{-1}}_{\left(a\right)}  \nonumber\\
\times&\left({{\mathbb{E}}^H}^k\left\{\mathbf{r}\right\}+{\delta }_{1}{\mathbf{W}}^{k+1}-{\mathbf{Z}}^k_1-{{\mathbb{E}}^H}^k{\mathbb{E}}^k\left\{{\mathbf{S}}^k\right\}\right),
\end{align}
\begin{align}\label{EQ9}
{\widehat{\mathbf{S}}}^{k+1}=&\underbrace{{\left({{\mathbb{E}}^H}^k{\mathbb{E}}^k+{\delta }_{2}\mathbf{I}\right)}^{-1}}_{\left(b\right)}
\nonumber\\ &
\times\left({{\mathbb{E}}^H}^k\left\{\mathbf{r}\right\}+{\delta }_{2}{\mathbf{Q}}^{k+1}-{\mathbf{Z}}^k_2-{{\mathbb{E}}^H}^k{\mathbb{E}}^k\left\{{\mathbf{L}}^k\right\}\right),
\end{align}
where ${\mathbb{E}}^H\left\{\cdot\right\}\triangleq\mathcal{P}\left\{{\mathbf{H}}^H{\mathbf{\Theta}}^T{\mathbf{P }}^H\left(\cdot\right)\right\}$,  ${\mathbf{Z}}_1$, ${\mathbf{Z}}_2$ are Lagrange multipliers, ${\delta }_1$ and ${\delta }_2$ denote for penalty parameters \cite{moradikia}.
The penalty
parameters can be set to the specific value ${\delta }_{1,2} = (nK)/(4 \Vert \mathbf{F} \Vert_1)$ as suggested in
\cite{Candes,Yuan} or updated dynamically according to \cite{Lin}.
${\mathrm{Prox}}_{\rho {\left\|.\right\|}_1}\left\{\cdot\right\}$ and ${\mathrm{Prox}}_{\rho {\left\|\cdot\right\|}_*}\left\{\cdot\right\}$ with parameter $\rho $, denote the soft-thresholding and the SVT operators \cite{moradikia}, respectively. 

Similarly, to estimate the vector $\mathbf{p }$ in \eqref{EQ5} we can again resort to the conventional ADMM. To do so, one can define new auxiliary variables $({\mathbf{w}}_{1},{\mathbf{w}}_{2})$, followed by minimizing the corresponding augmented function over each parameter separately. The solution at the $\left(k+1\right)$th iteration is thus obtained as:
\begin{align}\label{EQ10}
{\widehat{\mathbf{w}}}^{k+1}_1=\frac{{\rho }_1}{{\rho }_1+2{\lambda }_{\phi }}\left({\mathbf{p }}^k+{\rho }^{-1}_1{\boldsymbol{\mathrm{\Upsilon }}}^k_1\right),
\end{align}
\begin{align}\label{EQ11}
{\widehat{\mathbf{w}}}^{k+1}_2={\mathrm{Prox}}_{({-2{\lambda }_{\phi }}/{{\rho }_2}){\left\|.\right\|}_1}\left({\mathbf{p }}^k+{\rho }^{-1}_2{\boldsymbol{\mathrm{\Upsilon }}}^k_2\right),
\end{align}
\begin{align}\label{EQ12}
{\widehat{\mathbf{p}}}^{k+1}=&\underbrace{{\left({{\boldsymbol{\mathfrak{T}}}^H}^k{\boldsymbol{\mathfrak{T}}}^k+\left({\rho }_1+{\rho }_2\right)\mathbf{I}\right)}^{-1}}_{\left(c\right)}
\nonumber\\ &
\times\left({{\boldsymbol{\mathfrak{T}}}^H}^k\mathbf{r}+{\rho }_1{\mathbf{w}}^{k +1}_1+{\rho }_2{\mathbf{w}}^{k+1}_{2}-(\boldsymbol{\mathrm{\Upsilon}}^k_1+{\boldsymbol{\mathrm{\Upsilon }}}^k_2)\right),
\end{align}
where ${\{{\rho }_i\}}^2_{i=1}$, ${{\{\boldsymbol{\mathrm{\Upsilon }}_i\}}}^2_{i=1}$, and ${\lambda }_{\phi }$ represent the penalty parameters, Lagrange multipliers, and the regularization parameter, respectively. The estimated $\widehat{\mathbf{p}}$ is then used to update \eqref{EQ4} and the procedure continues unless the stopping criterion is satisfied.

\section{Proposed Fast ADMM-Based Solution}\label{propsed}
The brunt of the computational burden associated with the conventional ADMM steps \eqref{EQ6}-\eqref{EQ12} stems from the following:
 \renewcommand{\labelenumi}{\roman{enumi})}
\begin{enumerate}[label=(\roman*)]
	 \item The matrix inversion operations (a), (b), (c) in \eqref{EQ8}, \eqref{EQ9}, and \eqref{EQ12}, respectively. 
	
	\item Stacking the received echo and the unknown image into the large vectors $\mathbf{r}$ and $\mathbf{f}$, as well as the SAR kernel $\mathbf{H}$ with $N_a N_r \times I$ elements, which is not efficient in terms of increased storage and computational costs compared to the 2-D signal model which requires ($N_a \sqrt{I} +N_r \sqrt{I}$) parameters altogether\cite{Qiu}.
		
	\item The NP-hard problem of optimizing a quadratic form while adhering to the unimodular vector constraint as seen in Eq. \eqref{EQ5}. 
\end{enumerate}
In what follows, we have presented our alternative solutions in three different sub-sections to deal with each of abovementioned concerns (i)-(iii):
\subsection{Proposed Solution for Concern (i)}\label{sec3a}
By taking advantage of the matrix inversion lemma, we can replace \eqref{EQ8}, \eqref{EQ9}, and \eqref{EQ12}, with the following efficient versions in which costly matrix inversions are avoided:
\begin{align}\label{EQ13}
{\widehat{\mathbf{L}}}^{k+1}=&\left({\mathbf{W}}^{k+1}-{\ }{\delta }^{-1}_1{\mathbf{Z}}^k_1\right) \nonumber
  \\ &
-\frac{1}{1+{\delta }_1\ }\underbrace{{\mathbb{E}}^H\left\{\mathbb{E}\left\{{\mathbf{W}}^{k+1}-{\delta }^{-1}_1{\mathbf{Z}}^k_1+\mathbf{S}\right\}-\mathbf{r}\right\}}_{\left(d\right)},
\end{align}
\begin{align}\label{EQ14}
{\widehat{\mathbf{S}}}^{k+1}=&\left({\mathbf{Q}}^{k+1}-{\delta }^{-1}_2{\mathbf{Z}}^k_2\right)
\nonumber
\\ &
-\frac{1}{1+{\delta }_2}\underbrace{{\mathbb{E}}^H\left\{\mathbb{E}\left\{{\mathbf{Q}}^{k+1}-{\delta }^{-1}_2{\mathbf{Z}}^k_2+{\mathbf{L}}^k\right\}-\mathbf{r}\right\}}_{\left(e\right)},
\end{align}
\begin{align}\label{EQ15}
{\widehat{\mathbf{p}}}^{k+1}={\boldsymbol{\mathrm{\Lambda }}}_{\widetilde{\mathbf{h}}}\times\left({{\boldsymbol{\mathfrak{T}}}^H}^k\boldsymbol{\mathrm{r}}+{\rho }_1{\boldsymbol{\mathrm{w}}}^{k+1}_1+{\rho }_2{\boldsymbol{\mathrm{w}}}^{k+1}_2-{\boldsymbol{\mathrm{\Upsilon }}}^k_1-{\boldsymbol{\Upsilon }}^k_2\right),
\end{align}
where ${\boldsymbol{\mathrm{\Lambda }}}_{\widetilde{\mathbf{h}}}\triangleq \mathrm{diag}\left\{\frac{1}{{\tilde{h}}_1},\frac{1}{{\tilde{h}}_2},\dots ,\frac{1}{{\tilde{h}}_M}\right\}$ with ${\tilde{h}}_i$ representing the $i$th element of  $\widetilde{\mathbf{h}}\triangleq {\left(\boldsymbol{\mathrm{\Theta }}\mathbf{H}\mathbf{f}\right)}^{*}\odot \left(\boldsymbol{\mathrm{\Theta }}\mathbf{H}\mathbf{f}\right)+\left({\rho }_1+{\rho }_2\right)\mathbf{1}$. A proof of the above is provided in Appendix A.

\subsection{Proposed Solution for Concern (ii)}\label{sec3b}

\begin{figure*}
	\begin{align}\label{EQ17}
	\widehat{\mathbf{L}}=\left({\mathbf{W}}^{k+1}-{\delta }^{-1}_1{\mathbf{Z}}^k_1\right)
	-\frac{1}{1+{\delta }_1\ }\mathcal{G}\left\{{\mathbf{F}}^H_{a}{\mathbf{\Theta}}_a^T{\mathbf{\Phi }}^H\left[{\boldsymbol{\mathrm{\Phi }}\boldsymbol{\Theta }}_{a}{\mathbf{F}}_{a}{\mathcal{G}}^{-1}\left\{{\mathbf{W}}^{k+1}-{\delta }^{-1}_1{\mathbf{Z}}^k_1+{\mathbf{S}}^k\right\}{\mathbf{F}}^{T}_{r}{\boldsymbol{\mathrm{\Theta }}}^{T}_{r}-\mathbf{R}\right]{\boldsymbol{\mathrm{\Theta }}}_{r}{\mathbf{F}}^{*}_{r}\right\},
	\end{align}
	\begin{align}\label{EQ18}
	\widehat{\mathbf{S}}=\left({\mathbf{Q}}^{k+1}-{\delta }^{-1}_2{\mathbf{Z}}^k_2\right)-\frac{1}{1+{\delta }_2}\mathcal{G}\left\{{\mathbf{F}}^H_{a}{\mathbf{\Theta}}_a^T{\mathbf{\Phi }}^H\left[{\boldsymbol{\mathrm{\Phi }}\boldsymbol{\Theta }}_{a}{\mathbf{F}}_{a}{\mathcal{G}}^{-1}\left\{{\mathbf{Q}}^{k+1}-{\delta }^{-1}_2{\mathbf{Z}}^k_2+{\mathbf{L}}^k\right\}{\mathbf{F}}^{T}_{r}{\boldsymbol{\mathrm{\Theta }}}^{T}_{r}-\mathbf{R}\right]{\boldsymbol{\mathrm{\Theta }}}_{r}{\mathbf{F}}^{*}_{r}\right\}.
	\end{align}
	\hrulefill
\end{figure*}

To alleviate this concern, $\mathbf{r}$ and $\mathbf{f}$ have to be maintained in their original 2-D matrix forms $\mathbf{R}\in {\mathbb{C}}^{M_a\times M_r}$ and ${\mathcal{G}}^{-1}\left\{\mathbf{F}\right\}\in {\mathbb{C}}^{\sqrt{I}\times \sqrt{I}}$ rather than being reshaped into vectors. We note that $\mathbf{H}$ and $\boldsymbol{\mathrm{\Theta }}$ in \eqref{EQ2} can be decomposed into their azimuth and range components as  $\mathbf{H}={\mathbf{F}}_r\otimes {\mathbf{F}}_a$ and $\boldsymbol{\mathrm{\Theta}}={\boldsymbol{\mathrm{\Theta }}}_r\otimes {\boldsymbol{\mathrm{\Theta }}}_a$. The observation model \eqref{EQ2} may thus be recast as 
\begin{align}\label{EQ16}
\mathbf{R}=\boldsymbol{\Phi }{\boldsymbol{\Theta }}_{a}{\mathbf{F}}_{a} \mathbf{X}{\mathbf{F}}^T_r{\boldsymbol{\mathrm{\Theta }}}^T_r+\mathbf{Z},
\end{align}
where $\mathbf{X}={\mathcal{G}}^{-1}\{\mathbf{F}\}$ is the desired image of the scene, ${\mathbf{F}}_{\mathrm{a}}\in {\mathbb{C}}^{N_a\times \sqrt{I}}$ and ${\mathbf{F}}_{\mathrm{r}}\in{\mathbb{C}}^{N_r\times \sqrt{I}}$ denote the Fourier matrices in azimuth and range directions, respectively (as defined in Appendix A), $\mathbf{Z}\in {\mathbb{C}}^{M_a\times M_r}$ is the the undersampled Gaussian noise matrix, and ${\boldsymbol{\mathrm{\Theta }}}_{\mathrm{a}}\in{\mathbb{C}}^{M_a\times N_a}$ and ${\boldsymbol{\mathrm{\Theta }}}_r\in{\mathbb{C}}^{M_r\times N_r}$ denote the corresponding undersampling matrices.

Given the model in \eqref{EQ16}, we can formulate more efficient versions of \eqref{EQ13} and \eqref{EQ14}; the results of which are given by \eqref{EQ17} and \eqref{EQ18} shown at the top of this page. 
A proof of these relations is presented in Appendix B.

\subsection{Proposed Solution for Concern (iii)}
To deal with the optimization problem in \eqref{EQ5} more efficiently, we can resort to the recent advances in solving this class of problems \cite{Soltanalian1,Soltanalian2,Soltanalian3}. To do so, we can recast \eqref{EQ5} into the following equivalent unimodular quadratic program (UQP) form: 
\begin{align}\label{EQ19}
\widehat{\mathbf{p}}={\mathrm{arg} {\mathop{\mathrm{max}}_{\mathbf{p}} {\widetilde{\mathbf{p}}}^H\mathbf{U}\widetilde{\mathbf{p }}\ }\ }
\ \ \mathrm{s.t.}\ \ \ \ \left|{\left(\widetilde{\mathbf{p}}\right)}_i\right|=1,\ \ \ \ \ \forall i
\end{align}
where $\widetilde{\mathbf{p }}\triangleq {\left[\mathbf{p},1\right]}^T$ and the positive definite matrix $\mathbf{U}$ is formed as
$\mathbf{U}\triangleq\mu \mathbf{I}-\tilde{\boldsymbol{\mathfrak{T}}}$,  with 
\begin{equation}\label{EQ21}
\tilde{\boldsymbol{\mathfrak{T}}}\triangleq \left[ \begin{array}{cc}
{\boldsymbol{\mathfrak{T}}}^H\boldsymbol{\mathfrak{T}} & -{\boldsymbol{\mathfrak{T}}}^H\boldsymbol{\mathrm{r}} \\ 
-{\boldsymbol{\mathrm{r}}}^H\boldsymbol{\mathfrak{T}} & 0 \end{array}
\right]
\end{equation}
 and $\mu $ being an arbitrary real-valued constant larger than the maximum eigenvalue of $\boldsymbol{\mathfrak{T}}$ \cite{Soltanalian1,Soltanalian2,Soltanalian3}. Note that \eqref{EQ19} is
NP-hard in general. To approximate the solutions to \eqref{EQ19}, a computationally efficient approach was introduced in \cite{Soltanalian1,Soltanalian2,Soltanalian3} where $\mathbf{p}$ is updated through the following nearest vector problems with convergence guarantees:
\begin{align}\label{EQ22}
\widehat{\mathbf{p}}={\mathrm{arg} {\mathop{\mathrm{min}}_{\mathbf{p}} {\left\|{\widetilde{\mathbf{p}}}^{t+1}-\mathbf{U}{\widetilde{\mathbf{p}}}^{t}\right\|}_2\ }\ },\ \mathrm{s.t.} \ \left|{\left(\widetilde{\mathbf{p}}^{t+1}\right)}_i\right|=1,\ \forall i.
\end{align}
The closed-form solution of \eqref{EQ22} is simply obtained as a set of \textit{power} \textit{method-like} iterations, given by
\begin{align}\label{EQ23}
{\widetilde{\mathbf{p}}}^{t+1}=e^{j\mathrm{arg}\left(\mathbf{U}{\widetilde{\mathbf{p}}}^{t}\right)}
\end{align}

In fact, using the recursive formula of \eqref{EQ23}, an inner loop is added to each ADMM iteration whose $(t+1)$th iteration consists of monotonically boosting the criterion of \eqref{EQ22}. The proof of the monotonically increasing behaviour of the criterion in \eqref{EQ22} through the \textit{power} \textit{method-like} iterations \eqref{EQ23} is presented in Appendix C. Therefore, we can simply surrogate \eqref{EQ15} by the recursive formula of \eqref{EQ23}. The inner loop continues unless a maximum tolerable iteration is reached. 

Based on the above solutions, the proposed fast ADMM-based scheme, is presented in Algorithm \ref{Alg1}. 

\begin{algorithm}
	\caption{Fast ADMM-based LRSD-SAR Imaging}
	\label{Alg1}
	\begin{algorithmic}[1]
	\State \textbf{Input:}
	${\delta }_{1}, {\delta }_{2}$, ${\rho }_{1}\,{\rho }_{2}$, $\eth $, ${\alpha }_x$, ${\lambda }_L,{\lambda }_S $;
	\State \textbf{Initialization:}
	$\mathbf{F}^0={\mathbf{L}}^0={\mathcal{P}}\left( \left(\mathbf{\Theta}\mathbf{H}\right)^{H} \mathbf{r}\right)$, ${\mathbf{S}}^{0}=\boldsymbol{0}$, $\boldsymbol{\Phi }=\mathbf{I}$, $\left\{\mathbf{Z}_i^{0}\right\}^2_{i=1} = \left\{\mathbf{\Upsilon}_i^{0}\right\}^2_{i=1}=\mathbf{0}$, $k=0$;
	\While{${\left \Vert{\left|\mathbf{f}\right|}^{k+1}-{\left|\boldsymbol{\mathrm{f}}\right|}^k\right \Vert}_2/\ {\left \Vert{\left|\boldsymbol{\mathrm{f}}\right|}^k\right \Vert}_{\mathrm{2}}\mathrm{<}{\alpha }_x$ } \textbf{(I)-(IV)}
	\State    \textbf{I:}  \textbf{Reconstruction of} $\mathbf{f}$
	\State        $\ t=0$;
	\State         Calculate \eqref{EQ6}, \eqref{EQ7}, \eqref{EQ17}, \eqref{EQ18};
	\State         ${\mathbf{f}}^{k+1}\leftarrow {\mathcal{P}}^{-1}\left\{{\mathbf{L}}^{k+1}+{\mathbf{S}}^{k+1}\right\}$;
	\State     \textbf{II: Estimation of }$\mathbf{p }$
	\While	{$\left \Vert\mathrm{arg} \left(  \mathbf{p }^{t+1}-\mathbf{p }^t \right)  \right \Vert_2/ \left \Vert\mathrm{arg} \left( \mathbf{p }^t \right) \right \Vert_2 <\eth $} \State                 \textbf{\textit{(i):}}  Calculate \eqref{EQ23}
	\State             \textbf{\textit{(ii):
	}}  Extract $\mathbf{p }$ from $\widetilde{\mathbf{p }}$ by change of the variable
\EndWhile
\State  \textbf{   III: Model Matrices Update}
	\State         ${\mathbb{E}}^{k+1}\left\{.\right\}=\boldsymbol{\mathrm{\Theta }}{\boldsymbol{\mathrm{P}}}^{k+1}\boldsymbol{\mathrm{H}}{\mathcal{P}}^{-1}\left\{.\right\}$,
	\State ${\boldsymbol{\mathfrak{T}}}^{k+1}=\mathrm{diag}\left\{\boldsymbol{\mathrm{\Theta }}\boldsymbol{\mathrm{H}}{\boldsymbol{\mathrm{f}}}^{k+1}\right\}$, 
	\State Building ${\boldsymbol{\mathrm{\Phi }}}^{k+1}$ from ${\mathbf{p }}^{k+1}$
	\State \textbf{  IV: Lagrange Multipliers Update }
	\State  ${\boldsymbol{\mathrm{Z}}}^{k+1}_1={\boldsymbol{\mathrm{Z}}}^k_1+{\delta }^k_1\left({\mathbf{L}}^{k+1}-{\mathbf{W}}^{k+1}\right)$;
	\State  ${\mathbf{Z}}^{k+1}_2={\boldsymbol{\mathrm{Z}}}^k_2+{\delta }^k_2\left({\mathbf{S}}^{k+1}-{\mathbf{Q}}^{k+1}\right)$
\EndWhile
	\State Output: $\mathbf{S}$, $\mathbf{L}$. \end{algorithmic}
\end{algorithm}

\subsection{Accelerating the Autofocusing Algorithm}
Due to the large sizes of $\mathbf{U}$ and $\mathbf{p}$, the computational cost of \eqref{EQ23} is considerable. According to \eqref{Eq-phi}, however, in the vector $\mathbf{p}$ we only have $M_a$ distinct phase error components which are repeated $M_r$ times. This observation can be used to simplify \eqref{EQ23}. In fact, by combining \eqref{EQ5}, \eqref{EQ21} and \eqref{EQ23}, and some trivial algebraic manipulations, we obtain,
\begin{equation}\label{eq23}
\mathrm{arg}\left(\mathbf{p}^{t+1}\right)=\left[ \mu \mathbf{1}- \left(\boldsymbol{\Theta }\mathbf{H}\mathbf{f}\right)^*\odot\left(\boldsymbol{\Theta }\mathbf{H}\mathbf{f}\right)\right]\odot\mathbf{p}^{t}+\left(\boldsymbol{\Theta }\mathbf{H}\mathbf{f}\right)^*\odot \mathbf{r}
\end{equation}
Now, assume that the range compression has been performed. Consequently, we can further simplify \eqref{eq23} by substituting the equivalent matrix representation model \eqref{EQ16} and averaging the estimated phase error over various range cells, leading to:
\begin{align}\label{eq24}
\phi_i^{k+1}= & \mathrm{arg} \left\{\frac{1}{M_r}
\sum_{j=1}^{M_r}\left[ \left( \mu-  \mathbf{Y}_{i,j}^* \mathbf{Y}_{i,j}  \right) e^{j\phi_i^k}+ \mathbf{Y}_{i,j}^* \tilde{\mathbf{R}}_{i,j}  \right]   \right\}, 
\nonumber \\
& \mathrm{for} \ \ \  1 \leq i \leq  M_a,
\end{align}
where $\mathbf{Y}=\mathbf{\Theta}_a\mathbf{F}_a \mathbf{X}$, and $\tilde{\mathbf{R}}=\mathbf{R}\mathbf{\Theta}_r\mathbf{F}^{*}_r $ is the range compressed signal. 
We update the PRPE in the main loop, in which the SAR image is obtained, and thus, the inner loop is omitted. Moreover, \eqref{eq24} can be written efficiently in the matrix form as
\begin{align}\label{eq25}
\boldsymbol{\phi}^{k+1}= & \mathrm{arg} 
\left\{ \mathrm{sum} \left(\boldsymbol{\Phi}^{k} ( \mu \mathbf{1}-\mathbf{Y}^* \odot \mathbf{Y} )+ \mathbf{Y}^* \odot \tilde{\mathbf{R}}
\right) \right\},
\end{align}
where $\boldsymbol{\phi}=[\phi_1,\phi_2,\cdots,\phi_{M_a}]^T$ and $\mathrm{sum}(\cdot)$
denotes a summation of
matrix elements in the column direction.
 The obtained simpified algorithm is presented in Algorithm \ref{Alg2}. For the purposes of simplicity, we denote our algorithm by ADMM-UQP in the rest of the paper.
\begin{algorithm}
	\caption{Simplified Fast ADMM-based LRSD-SAR Imaging}
	\label{Alg2}
	\begin{algorithmic}[1]
		\State \textbf{Input:}
		${\delta }_{1}$, ${\delta }_{2}$, ${\rho }_{1}$, ${\rho }_{2}$, ${\alpha }_x$, ${\lambda }_L,{\lambda }_S $;
		\State \textbf{Initialization:}
		${\mathbf{F}}^0={\mathbf{L}}^0={\mathcal{G}}\left(  \mathbf{F}_a^H \mathbf{\Theta}_a^T \mathbf{R} \mathbf{\Theta}_r \mathbf{F}_r^* \right)$, ${\mathbf{S}}^{0}=\boldsymbol{0}$, $\boldsymbol{\Phi }=\mathbf{I}$, $\left\{\mathbf{Z}_i^{0}\right\}^2_{i=1} = \left\{\mathbf{\Upsilon}_i^{0}\right\}^2_{i=1}=\mathbf{0}$, $k=0$;
		\While{${\left \Vert{\left|\boldsymbol{\mathrm{X}}\right|}^{k+1}-{\left|\boldsymbol{\mathrm{X}}\right|}^k\right \Vert}_2/\ {\left \Vert{\left|\boldsymbol{\mathrm{X}}\right|}^k\right \Vert}_{2}<{\alpha }_x$ } \textbf{(I)-(III)}
		\State    \textbf{I:}  \textbf{Reconstruction of} $\mathbf{X}$
		\State    Calculate       \eqref{EQ6}, \eqref{EQ7}, \eqref{EQ17}, \eqref{EQ18};
		\State         ${\mathbf{X}}^{k+1}\leftarrow {\mathcal{G}}^{-1}\left\{{\mathbf{L}}^{k+1}+{\mathbf{S}}^{k+1}\right\}$;
		\State     \textbf{II: Estimation of PRPE}
		 \State                 Calculate $\boldsymbol{\phi }^{k+1}$ from \eqref{eq25} 
		 \State                Update ${\boldsymbol{\Phi }}^{k+1}$
			\State \textbf{  III: Lagrange Multipliers Update }
			\State  ${\boldsymbol{\mathrm{Z}}}^{k+1}_1={\boldsymbol{\mathrm{Z}}}^k_1+{\delta }^k_1\left({\mathbf{L}}^{k+1}-{\mathbf{W}}^{k+1}\right)$;
			\State  ${\mathbf{Z}}^{k+1}_2={\boldsymbol{\mathrm{Z}}}^k_2+{\delta }^k_2\left({\mathbf{S}}^{k+1}-{\mathbf{Q}}^{k+1}\right)$
			\EndWhile
			\State Output: $\mathbf{S}$, $\mathbf{L}$. \end{algorithmic}
	\end{algorithm}

\subsection{Application in ISAR Imaging}
Due to the duality of SAR and ISAR modes and the similarity of the received signal models \cite{Chen}, we can develop the proposed method for ISAR imaging and autofocusing. However, we note that the SAR and ISAR scenes are fundamentally different. Since it is usually assumed that only one target exists in the ISAR scene (e.g. in the imaging of airborne or maritime targets), and the background only contains noise, the ISAR scene is inherently sparse but has no low-rank feature \cite{HashempourDSP}; i.e. $\mathbf{F}=\mathbf{S}$ and $\mathbf{L}=\mathbf{0}$. It is worth noting that we do not use the patch-based model for ISAR imaging, and can apply the algorithm directly to the scene image (i.e. $\mathbf{F}=\mathrm{Mat}\{f\}=\mathbf{X}$). Therefore, in order to use the proposed mehod for ISAR imaging, we only need to calculate \eqref{EQ7} and  \eqref{EQ18} in step 5 of Algorithm \ref{Alg2}. On the other hand, for ISAR autofocusing, we average the estimated phase in \eqref{eq24} for the range cells that contain the reflected echo of the target. This task is simply performed by utilizing a thershold in the ISAR image \cite{Zhang}.

\subsection{Comparison of the Computational Costs}
In this section, the computational complexity of the proposed method is compared with the conventional approaches in \cite{moradikia} and \cite{Video}. First, consider the LRSD algorithm for image formation. Conventional ADMM \cite{moradikia} basically consists of the operations \eqref{EQ6}-\eqref{EQ9}. The complexity of each step can be computed as follows. Equation \eqref{EQ6} actually relies on SVT. The largest computation cost of SVT is determined by the SVD, which is $O(nK \min (n,K))$ \cite{Golub}. Since the size of the vectorized form of the sliding window (i.e., $n$) is usually greater than the number of sliding windows
(i.e., $K$), the largest cost of SVT can be rewritten as $O(nK^2)$. Equation \eqref{EQ7} is simply governed by soft-thresholding which is mainly an element-wise multiplication by the cost of $O(nK)$. Without loss of generality, we set $I=N=M$ and $N_a=N_r=\sqrt I$, where $I$ is the size of
the vectorized form of the image. Thus, the most expensive operations in \eqref{EQ8} and \eqref{EQ9} is matrix inversion with a cost of $O(n^3)$ and matrix-vector multiplication which costs $O(I^2)$. However, by assuming $n^3>I^2$, the dominant cost is $O(n^3)$, whose computation may prove impractical for large scene sizes. Therefore the overall cost of conventional ADMM is $O(nK^2+nK+2n^3)$.
A Conjugate-Gradient (CG) algorithm is exploited in \cite{moradikia} to approximate the inverse.
However, applying the CG algorithm is also relatively inefficient.
In the proposed method, we first substitute  \eqref{EQ8} and \eqref{EQ9} with \eqref{EQ13} and \eqref{EQ14}, respectively, and then, the equivalent matrix representation is exploited that ultimately yields \eqref{EQ17} and \eqref{EQ18}.
The cost of computing \eqref{EQ13} grows as $O(I^2)$, while the same operation with \eqref{EQ17} costs $O(I^{3/2})$.
Thus, there is a $\sqrt I$ gain in the cost of the matrix version compared to the vector approach. The same gains are achieved for \eqref{EQ14} and \eqref{EQ18}. Consequently, the overal cost of the proposed method is $O(nK^2+nK+2I^{3/2})$. As a result, the image formation with the proposed method (i.e., the step 6 of Algorithm \ref{Alg1}) is performed more efficiently compared to the conventional approach. Specifically, the total gain in the cost of proposed method compared to the conventional one is $n^3/I^{3/2}$.

For autofocusing, using \eqref{EQ12} i.e. the method of \cite{Video}, will require computing a matrix inversion with the cost of $O(I^3)$ or applying the CG approximation algorithm \cite{Video}, which are both time consuming.  However, the matrix inversion is no longer required by employing \eqref{eq25} and the closed-form solution is obtained which only requires some element-wise operations with the complexity of $O(I^2)$. In the ADMM-SBL algorithm of \cite{admm-sbl}, the authors take advantage of  an entropy criterion and the Newton method for autofocusing. However, the calculations of the first and the second derivatives of entropy is time consuming compared to our proposed algorithm. Specifically, for each $\phi_i$, there are $O(I^2)$ element-wise operations, and for the total $I$ phase error elements, the complexity is $O(I^3)$. Consequently, our method has a gain of $I$ in the cost for autofocusing compared to \cite{admm-sbl} and \cite{Video}. 
The computation costs of the available
algorithms are summarized in Table \ref{table-complexity}.	

\begin{table}[!t]
	\small
	\renewcommand{\arraystretch}{1.3}
	\caption{Algorithms Comparison}
	\centering
	\begin{tabular}{c c c}
		\hline\hline
		Computation cost of:
		& Image Reconstruction &  Autofocusing\\
		\hline
		Method of \cite{moradikia,Video}  & $O(nK^2+nK+2n^3)$&   $O(I^3)$ \\
		Method of \cite{admm-sbl}   & $O(I^3)$   & $O(I^3)$		\\
		Proposed method  & $O(nK^2+nK+2I^{3/2})$  &  $O(I^2)$\\	
		\hline
	\end{tabular}
	\label{table-complexity}
\end{table}

\section{Experimental Results}\label{experimet}
In this section, several experiments on synthetic and real-world data are presented to evaluate the performance of the proposed algorithm. In particular, the range-Doppler algorithm (RDA), the ADMM-SBL \cite{admm-sbl}, and the conventional ADMM-based method (ADMM-Conv) \cite{moradikia,Video}, 
are compared with our ADMM-UQP algorithm.

\subsection{Synthetic Data Experiments}

\begin{figure} 
	\centering
	\subfloat[\label{fig.1a}]{%
		\includegraphics[width=0.5\linewidth]{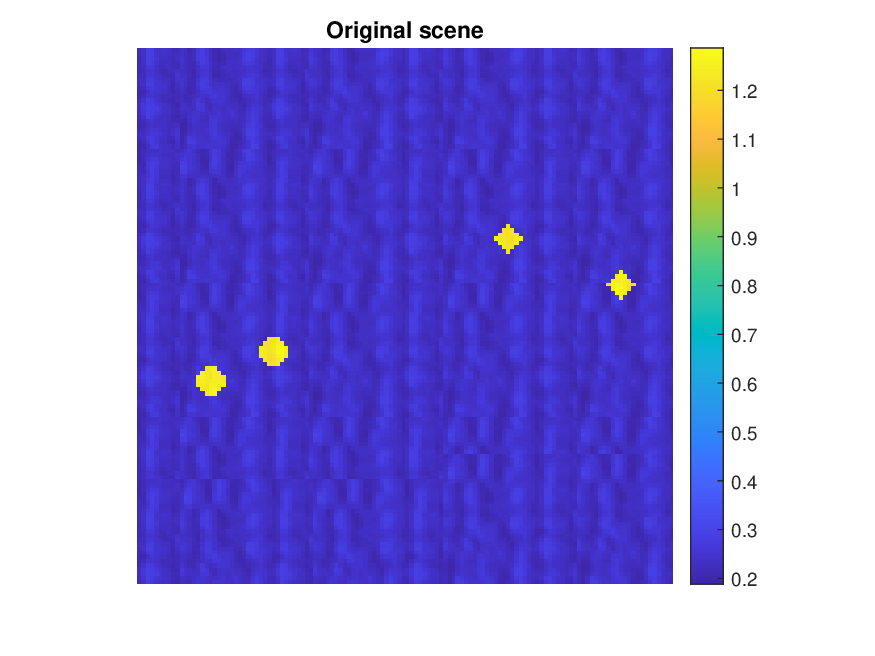}}
	\hfill
	\subfloat[\label{fig.1b}]{%
		\includegraphics[width=0.5\linewidth]{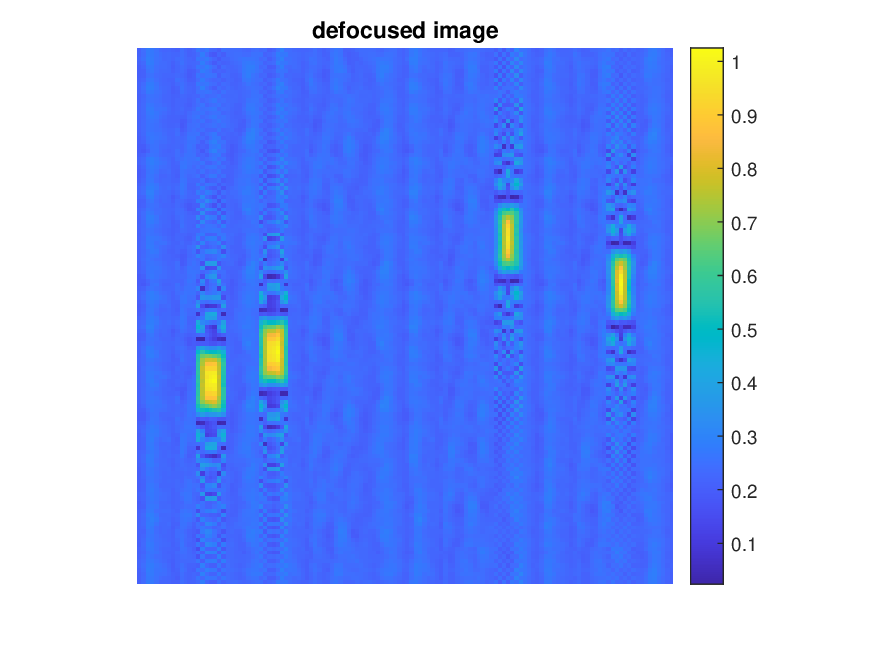}}
	\\[-2ex]
	\subfloat[\label{fig.1c}]{%
		\includegraphics[width=0.5\linewidth]{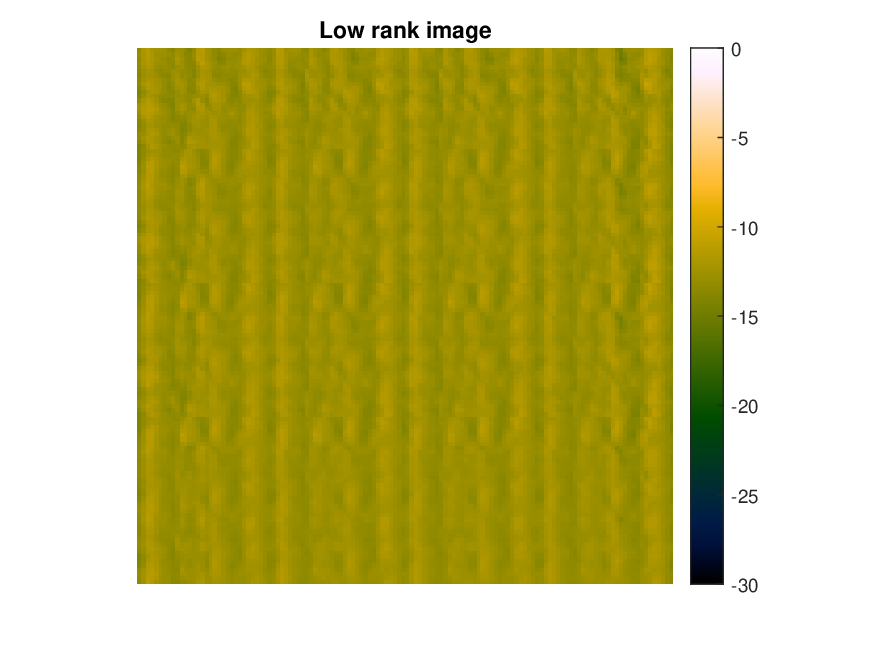}}\hfill
	\subfloat[\label{fig.1d}]{%
		\includegraphics[width=0.5\linewidth]{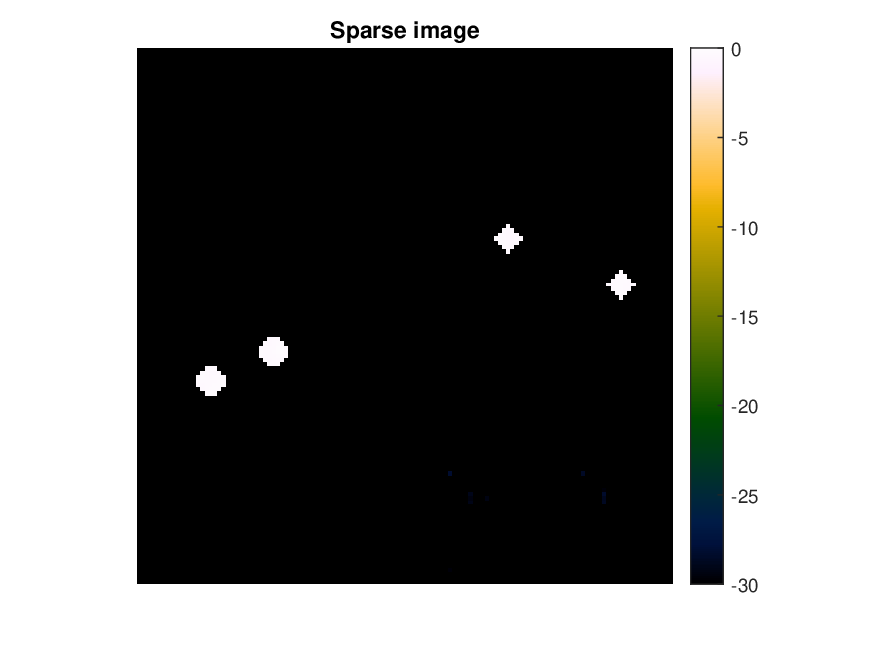}}
	\caption{Results of the proposed ADMM-UQP algorithm.
		\protect\subref{fig.1a} Original scene. \protect\subref{fig.1b} defocused image, \protect\subref{fig.1c} low-rank, 
		\protect\subref{fig.1d} and sparse images 
	}
	\label{fig1}		
\end{figure}

 We first consider a $128\times 128$ synthetic scene (Fig. \ref{fig.1a}) including a low-rank background and 4 point targets. 
Our simulation setting is based on the parameter values considered in \cite{moradikia}, unless otherwise stated. 
The PRPE is a $\pi/2$ quadratic phase error, and the observation noise, that follows a complex white Gaussian distribution, contaminates the phase history data with an SNR that is set to 10 dB. The length of the sliding window and the sliding step used in the patch-based operator $\mathcal{G}$ are set to 32 and 16, respectively. 
As observed in Fig. \ref{fig1}, the proposed algorithm, decomposes the low-rank and sparse components successfully. 
In order to compare the computational cost, the run-time (in seconds) of the proposed algorithm and the conventional method in \cite{moradikia} were computed over an Intel core i7, 4 GHz processor, for different image sizes, and the results are presented in Table \ref{table1}. 
It can be seen that the ADMM-UQP approach is at least three times faster than the conventional one in the performed experiments.

\begin{figure} 
	\centering
	\includegraphics[width=1\linewidth]{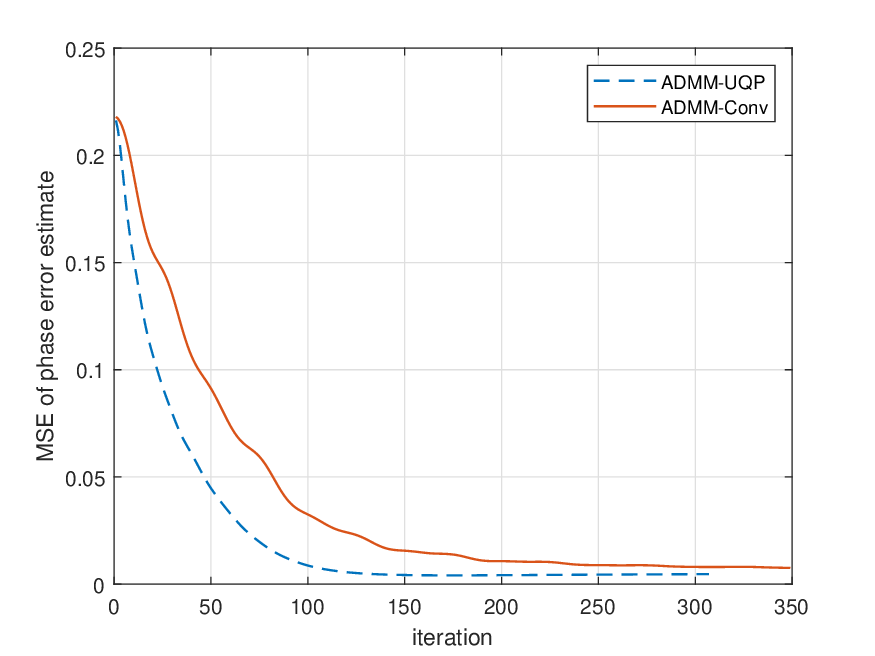}
	\caption{MSE of phase error estimation for a $\pi/2$ quadratic phase error. 
	}
	\label{fig2}		
\end{figure}

To investigate the convergence behavior, we have shown the behavior of the mean squared error (MSE) of phase error estimate  \cite{moradikia,Video} vs. the iteration number in Fig. \ref{fig2}. The MSE at the $k$th iteration is defined as,
\begin{equation}
\mathrm{MSE}_{\phi^k} = \frac{1}{M_a} \left \Vert \nabla \boldsymbol{\phi}^k \right \Vert_2^2.
\end{equation}
As witnessed, our approach converges in fewer iterations as compared to the approach in \cite{moradikia,Video}.

\begin{figure} 
	\centering
	\subfloat[\label{fig.3a}]{%
		\includegraphics[width=0.5\linewidth]{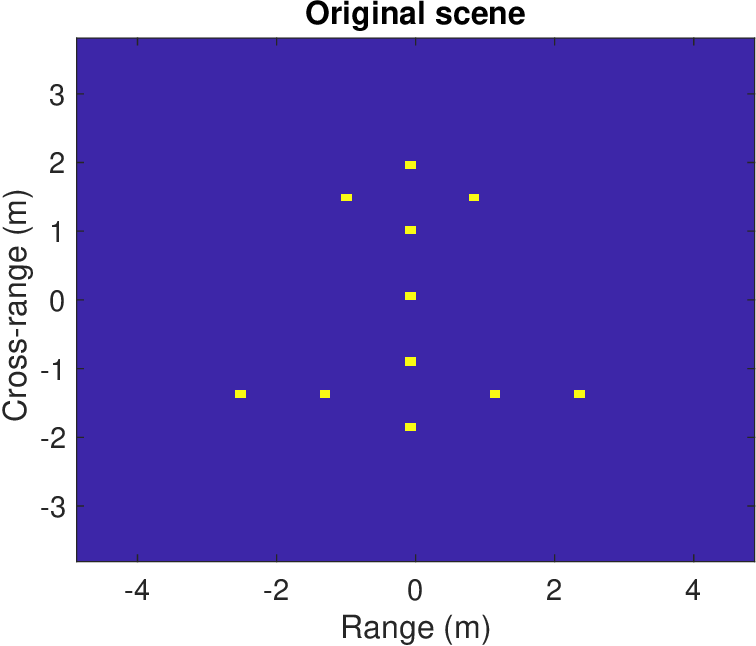}}
	\hfill
	\subfloat[\label{fig.3b}]{%
		\includegraphics[width=0.5\linewidth]{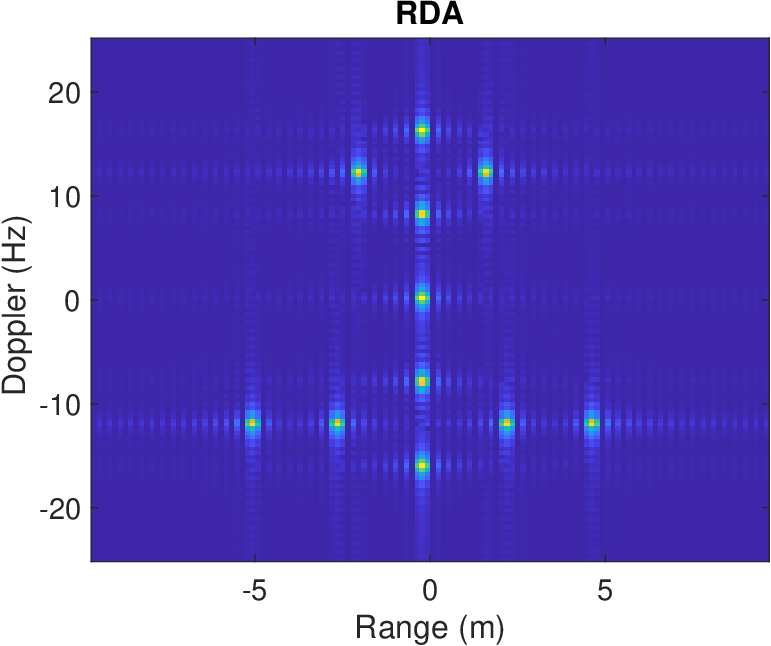}}
	\hfill
	\subfloat[\label{fig.3c}]{%
		\includegraphics[width=0.5\linewidth]{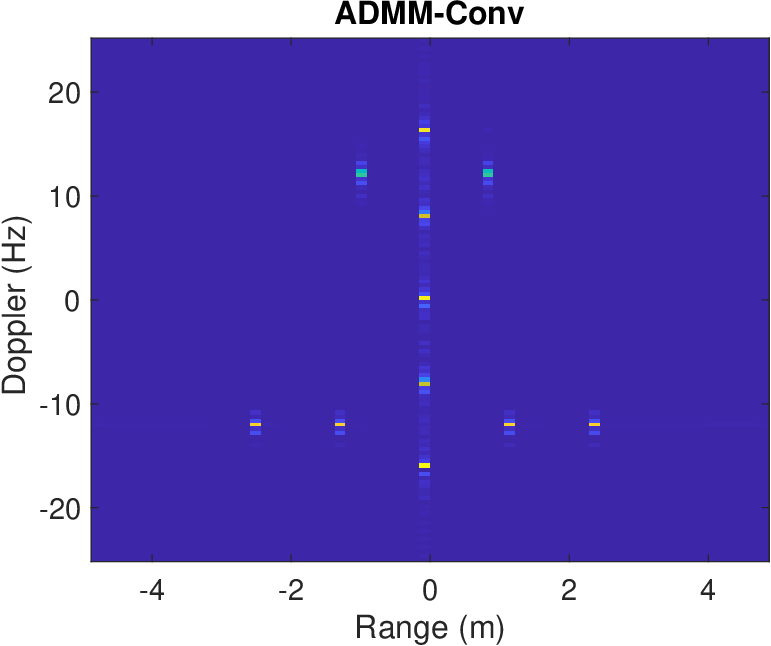}}
	\hfill
	\subfloat[\label{fig.3d}]{%
		\includegraphics[width=0.5\linewidth]{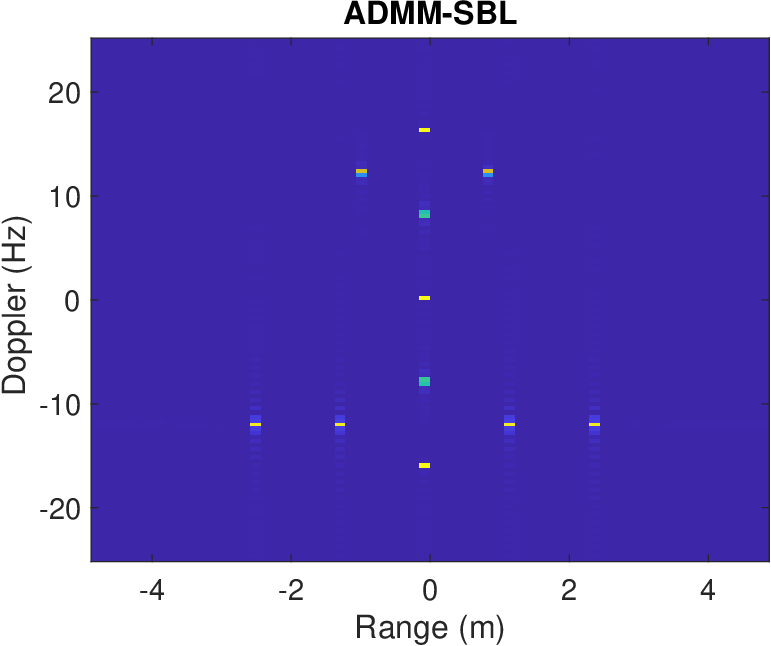}}
	\hfill
	\subfloat[\label{fig.3e}]{%
		\includegraphics[width=0.5\linewidth]{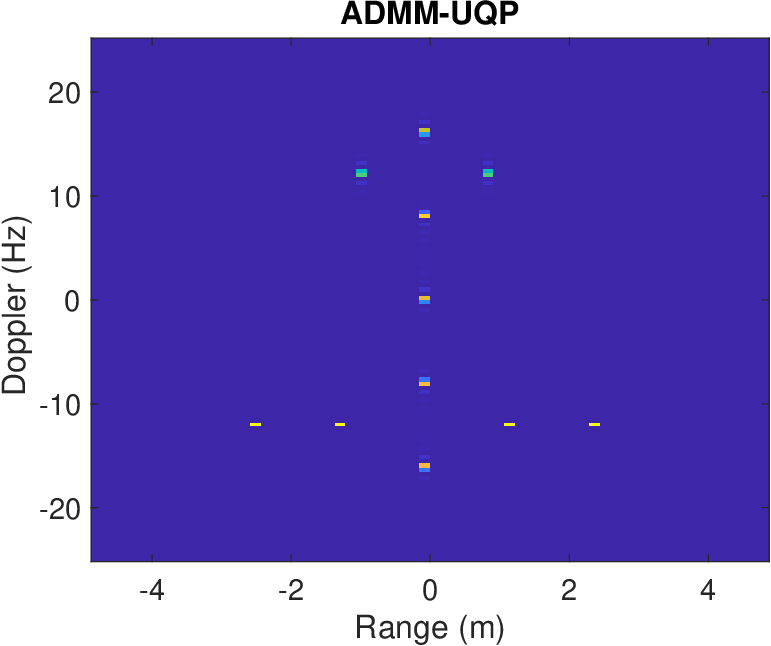}}
	\caption{ISAR imaging using synthetic data: \protect\subref{fig.3a} Original target scene, and the obtained ISAR images using \protect\subref{fig.3b} RDA, \protect\subref{fig.3c} ADMM-Conv, \protect\subref{fig.3d} ADMM-SBL, 
		\protect\subref{fig.3e} and ADMM-UQP. For all algorithms, the SNR is set to 10dB.}
		\label{fig3}
\end{figure}

In the next experiment, we consider  an ISAR scenario. The radar parameters are shown in Table \ref{table_radar_param}. The scene  size is $64 \times 64$, with 11 ideal point-like targets.
The RDA, ADMM-Conv, ADMM-SBL and ADMM-UQP algorithms have been applied to the simulated data to form the ISAR images. 
In order to have a fair comparison with ADMM-SBL, we apply range compression to the raw data. Therefore, the ISAR kernel is only comprised of the cross-range dictionary for  ADMM-Conv and ADMM-UQP. Moreover, for efficient autofocusing, we only use the range cells that contain the energy of the target in all autofocusing algorithms. 
The original target scene is displayed in Fig. \ref{fig.3a}. The ISAR images for SNR = 10dB are shown in Fig. \ref{fig3}. It is clear that, the image obtained by RDA suffers from sidelobes. However, the other three sparsity-driven methods achieve similar
images with high resolution and clean background. The quality of the image obtained from ADMM-UQP is slightly better than that of ADMM-Conv and ADMM-SBL, which
validate the effectiveness of the proposed approach.

\begin{table}[!t]
	\small
	\renewcommand{\arraystretch}{1.3}
	\caption{comparison of run-time for different image sizes}
	\centering
	\begin{tabular}{c c c c}
		\hline\hline
		Image size:
		& $64\times 64$  &  $128\times 128$ & $256\times 256$\\
		\hline
		ADMM-Conv \cite{moradikia}  & 1.6s&  5s & 25s\\
		ADMM-UQP & \textbf{0.4s}   & \textbf{1.2s} & \textbf{8s}		\\
		\hline
	\end{tabular}
	\label{table1}
\end{table}

\begin{table}[!t]
	\small
	\renewcommand{\arraystretch}{1.3}
	\caption{Radar parameters for synthetic data}
	\centering
	\begin{tabular}{c c c c c c}
		\hline\hline
		Carrier frequency $f_c$ & & & & & 10 GHz\\
		Bandwidth $B$ & & & & & 500 MHz\\
		Pulse Repetition Frequency $PRF$ & & & & &  50 Hz\\
		Number of range cells $N_r$ & & & & &  64\\
		Number of pulses $N_a$ & & & & &  64\\		
		\hline
	\end{tabular}
	\label{table_radar_param}
\end{table}  

We exploit the entropy values associated with the ISAR images to quantitatively compare the performance of different algorithms for three different SNR values as shown in Table \ref{table_Ent}.
The image entropy is defined as \cite{HashempourDSP},
\begin{equation}
I_x=- \sum_{n=1}^{N_a} \sum_{m=1}^{N_r} \dfrac{\vert \mathbf{X}_{n,m} \vert ^2}{E}
\log \dfrac{\vert \mathbf{X}_{n,m} \vert ^2}{E}
\end{equation}
where $E=\sum_{n=1}^{N_a} \sum_{m=1}^{N_r}\vert \mathbf{X}_{n,m} \vert ^2$ is the image energy.
Generally, a well-focused image has a low entropy value.
As it can be observed, the image entropy of the proposed algorithm in all cases is the lowest compared to the others. This appears to confirm the superiority and robustness of our algorithm.
Furthermore, in Table \ref{run_time}, the run-time of different methods are compared. Interestingly, the proposed ADMM-UQP is faster than the ADMM-Conv and ADMM-SBL algorithms in all SNR values. Specifically, due to the Bayesian parameters optimization and the calculations of the autofocusing algorithm in ADMM-SBL, the run time of this method is several folds larger than that of ADMM-UQP.
\begin{table}[!t]
	\small
	\renewcommand{\arraystretch}{1.3}
	\caption{Entropy of the ISAR image for different algorithms}
	\centering
	\begin{tabular}{c c c c}
		\hline\hline
		Algorithm & SNR = 10dB &  SNR = 0dB & SNR = -10dB \\
		\hline
		RDA  & 5.77 &  5.82  & 6.13\\
		ADMM-Conv \cite{moradikia}& 2.83   & 2.84 & 3.17\\
		ADMM-SBL  \cite{admm-sbl}& 2.64 &   2.75 & 3.01
		\\
		ADMM-UQP &  \textbf{2.60} & \textbf{2.61} & \textbf{2.65}
		\\
		\hline
		
	\end{tabular}
	\label{table_Ent}
\end{table}

\begin{table}[!t]
	\small
	\renewcommand{\arraystretch}{1.3}
	\caption{Run-time for different algorithms}
	\centering
	\begin{tabular}{c c c c}
		\hline\hline
		Algorithm & SNR = 10dB &  SNR = 0dB & SNR = -10dB \\
		\hline
		ADMM-Conv \cite{moradikia}& 7.35s   & 7.56s & 7.4s\\
		ADMM-SBL  \cite{admm-sbl}& 10.62s &   11.83s & 11.09s
		\\
		ADMM-UQP & \textbf{3.54s}  & \textbf{3.86s}  & \textbf{3.65s}
		\\
		\hline
		
	\end{tabular}
	\label{run_time}
\end{table}

\begin{figure*} 
	
	\medskip
	\begin{minipage}{0.03\textwidth}
  {\small (a)} \\  \\ \\ \\  \\ \\ \\
  \\ \\ {\small (b)} \\ \\ \\ \\ \\ \\ 
  \\ \\ \\{\small (c)}
	\end{minipage}\hfill
\begin{minipage}{0.24\textwidth}
	\includegraphics[width=\linewidth]{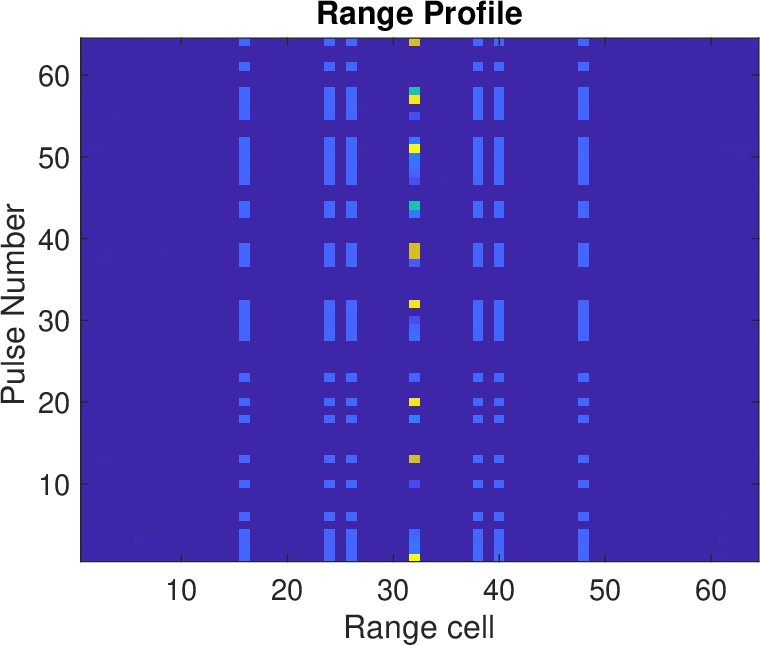}
	\includegraphics[width=\linewidth]{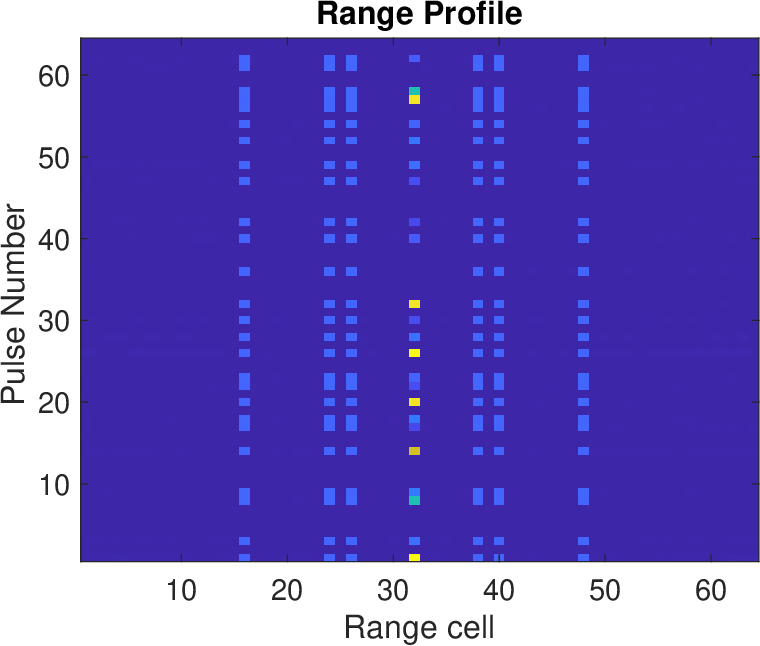}
	\includegraphics[width=\linewidth]{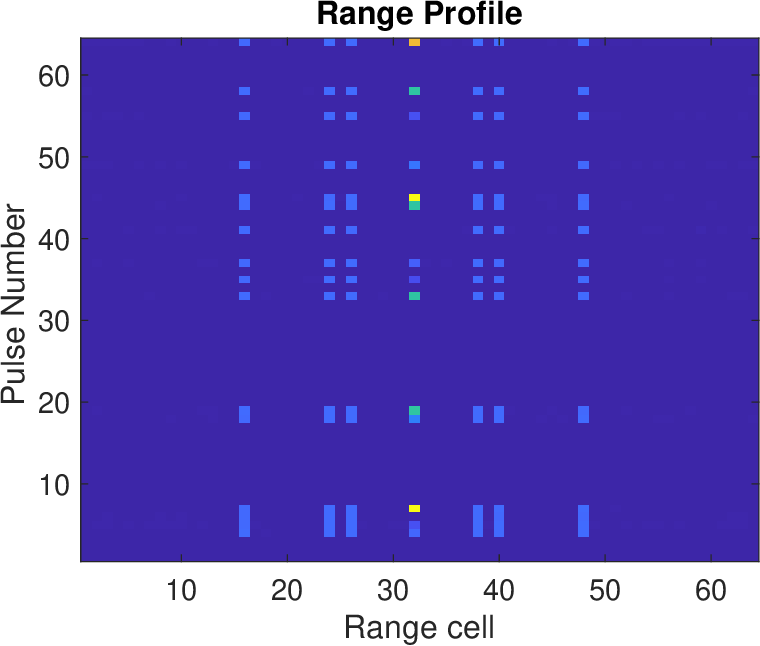}
\end{minipage}\hfill
\begin{minipage}{0.24\textwidth}
	\includegraphics[width=\linewidth]{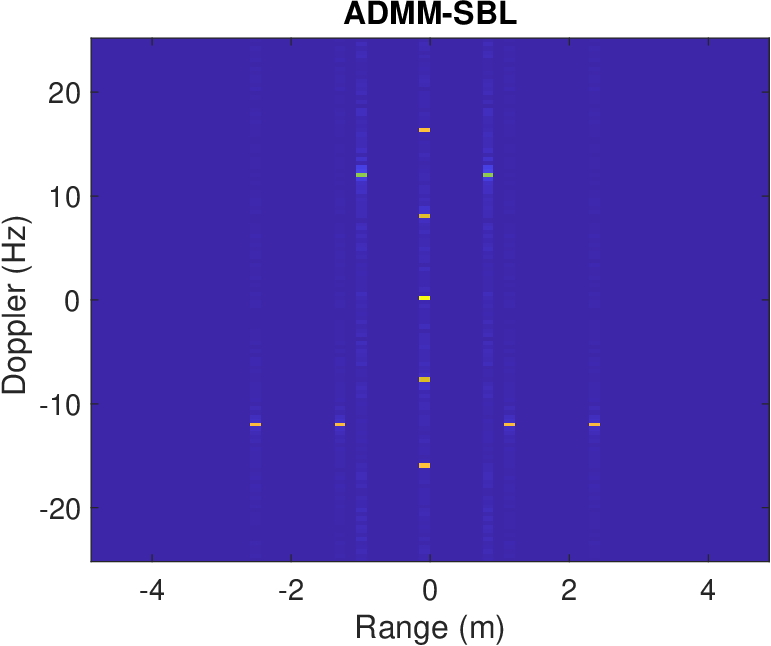}
	\includegraphics[width=\linewidth]{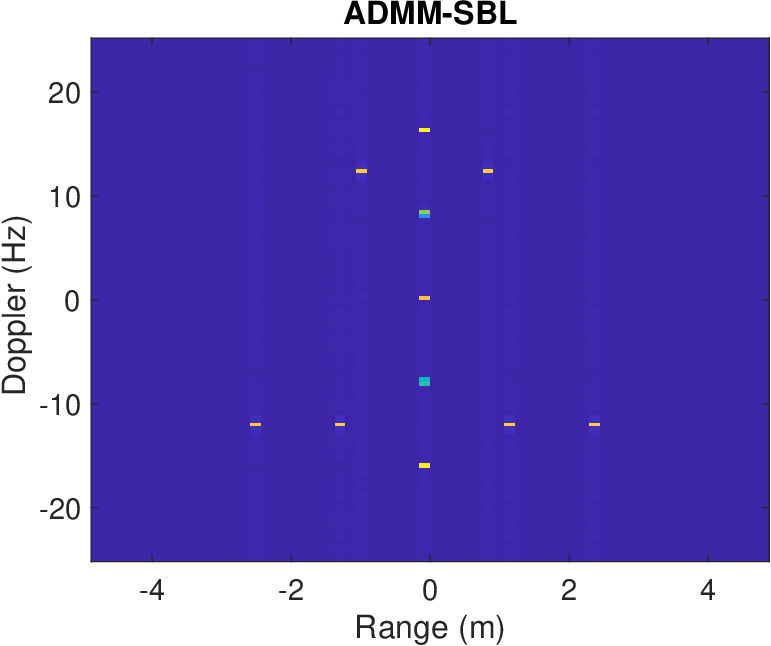}
	\includegraphics[width=\linewidth]{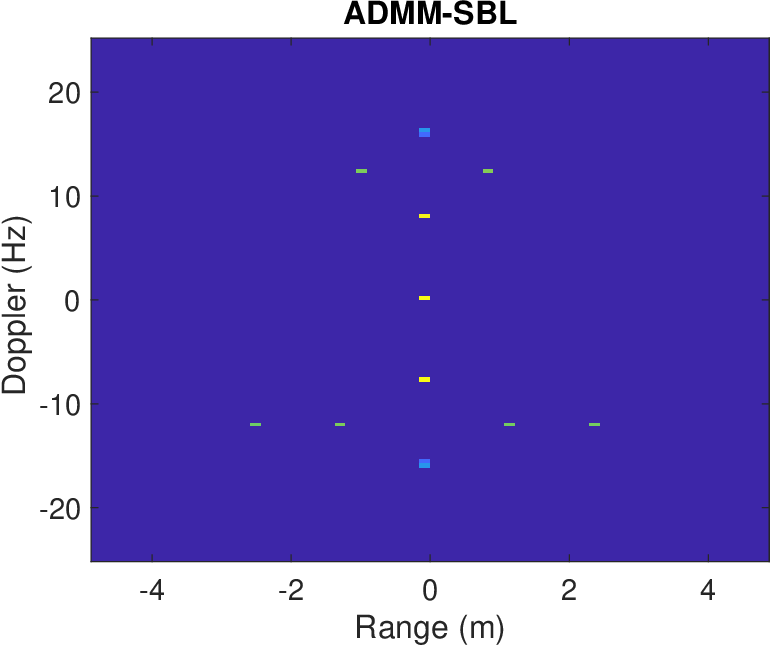}
\end{minipage}\hfill
\begin{minipage}{0.24\textwidth}
	\includegraphics[width=\linewidth]{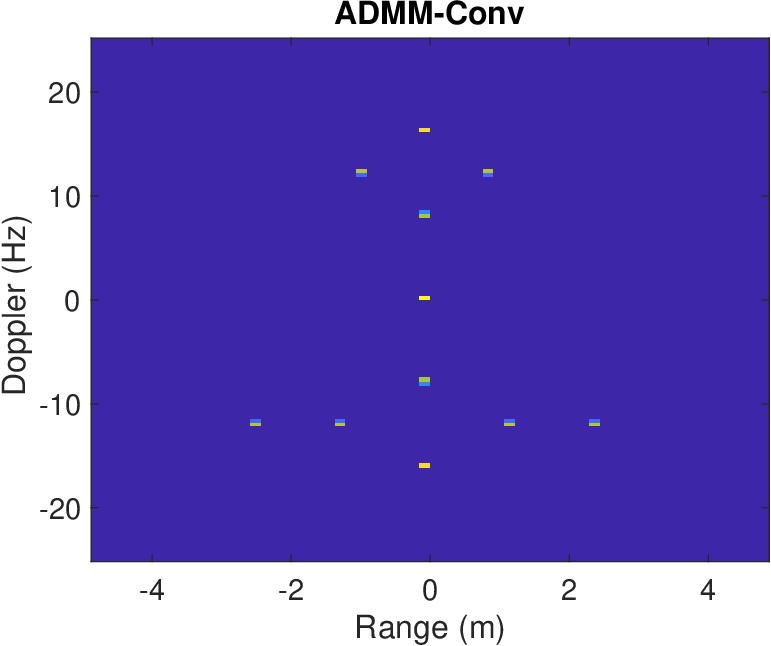}
	\includegraphics[width=\linewidth]{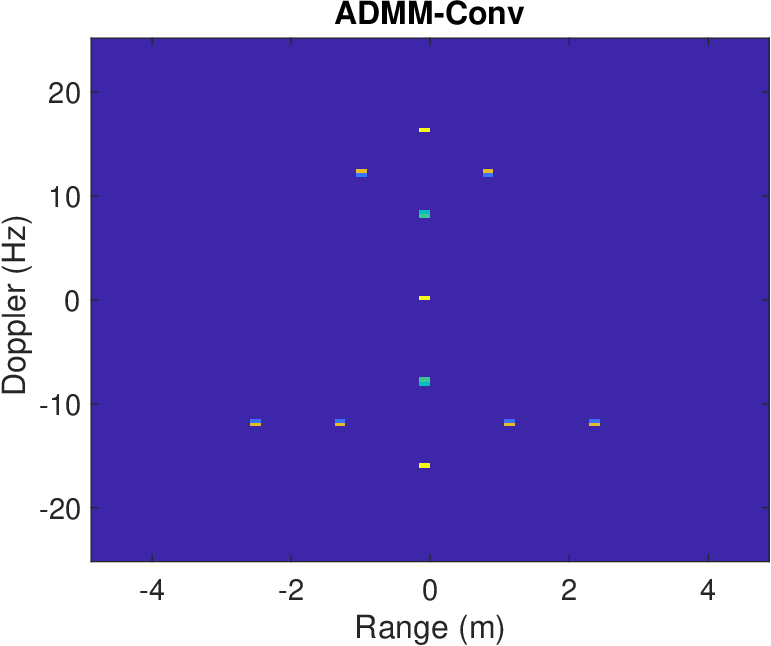}
	\includegraphics[width=\linewidth]{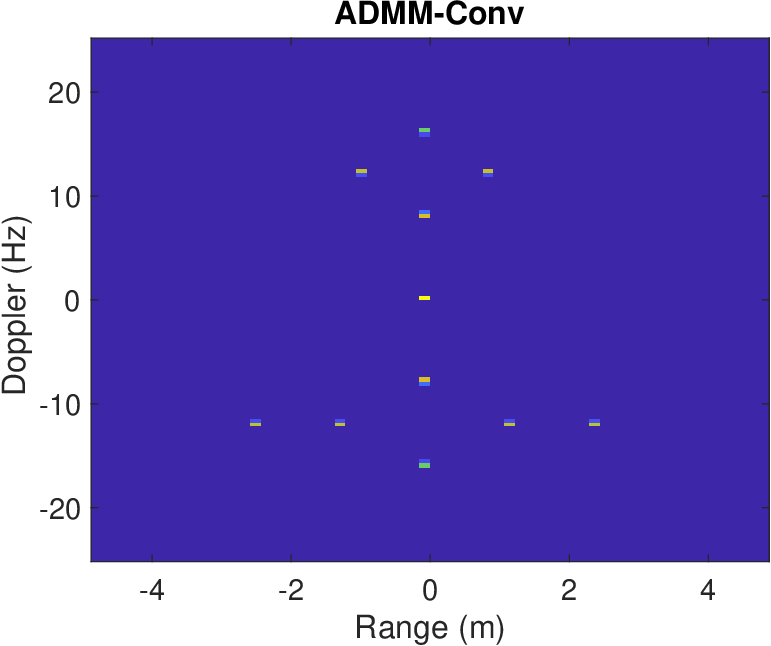}
\end{minipage}\hfill
\begin{minipage}{0.24\textwidth}
	\includegraphics[width=\linewidth]{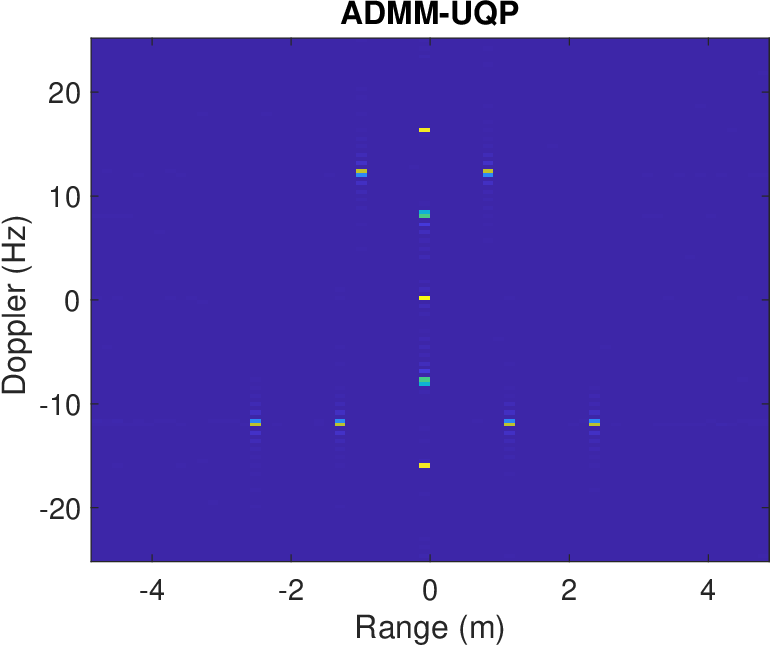}
	\includegraphics[width=\linewidth]{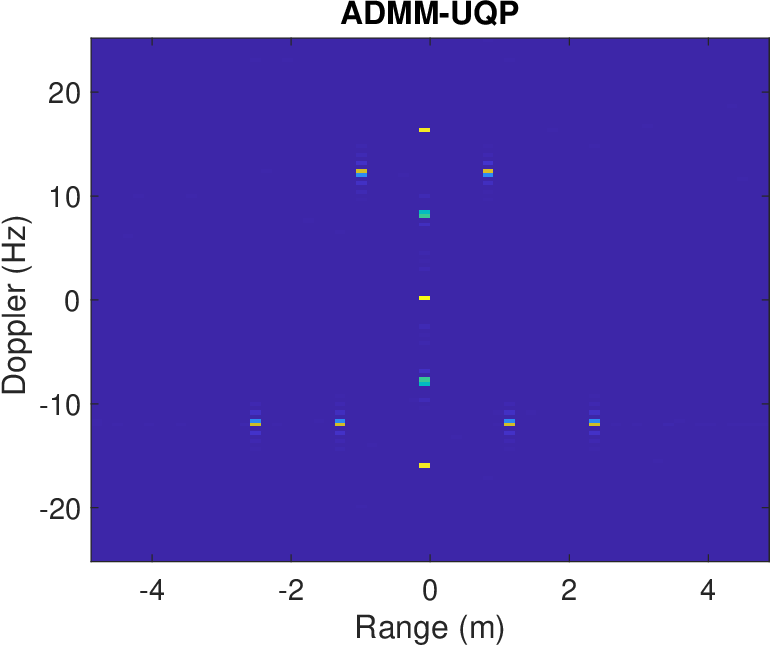}
	\includegraphics[width=\linewidth]{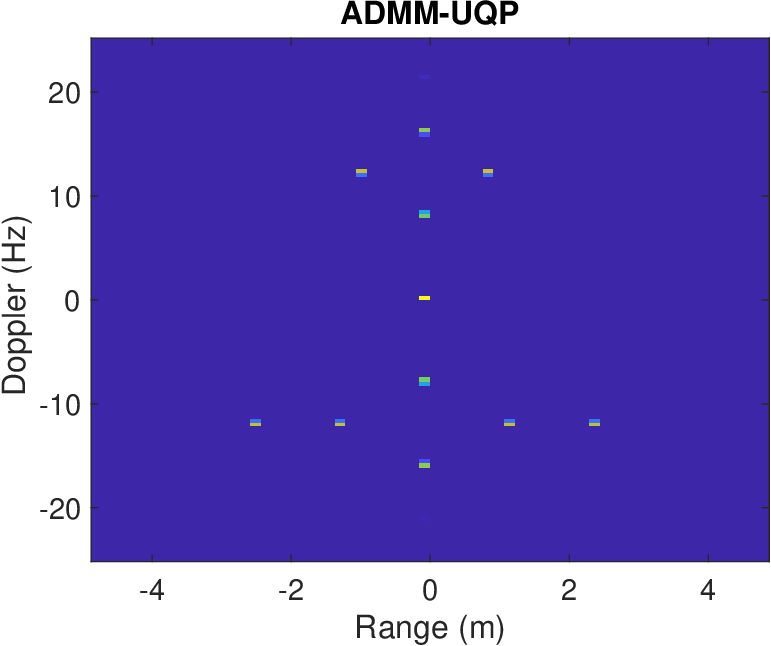}
\end{minipage}
\caption{Range profile and the ISAR images of the synthetic scene obtained from different methods using (a) 0.75, (b) 0.50, and (c) 0.25 of the measurements.}
\label{figg3} 
\end{figure*}

In the last simulation, we compare the algorithms when they are given undersampled data. Fig. \ref{figg3} demonstrates the range profile and ISAR images obtained by ADMM-SBL, ADMM-conv and ADMM-UQP with the undersampling ratios of 0.75, 0.50 and 0.25, respectively. It can be observed that all sparsity-driven algorithms reconstruct the original image successfully with small artifacts. However, the image artifacts of the proposed method appears less than that of the other algorithms. 

\subsection{Real-world Data Experiments}

\begin{figure} 
		\centering
	\subfloat[]{\label{fig4a}\includegraphics[width=0.5\linewidth]{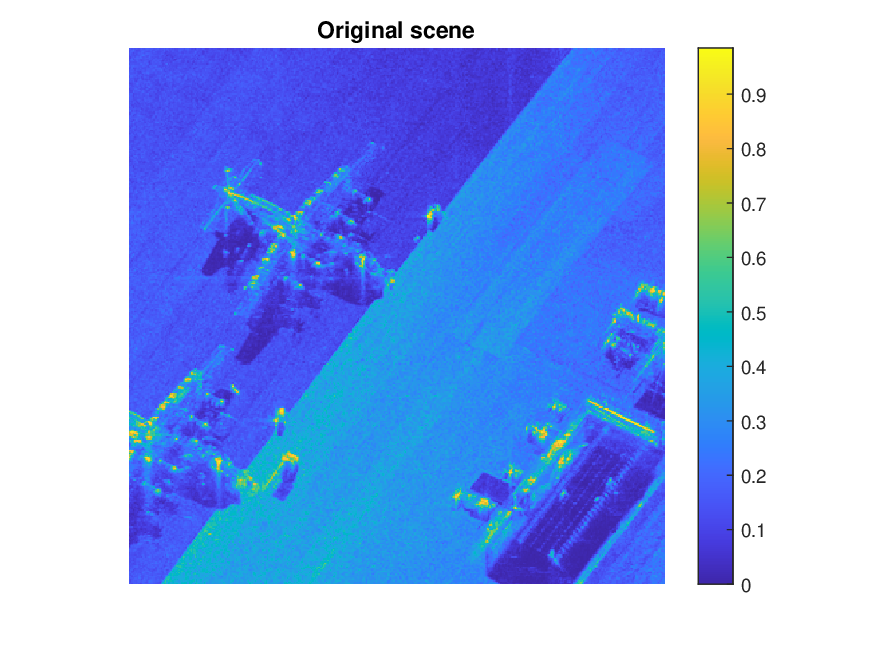}}\hfill 
	\subfloat[]{\label{fig4b}\includegraphics[width=0.5\linewidth]{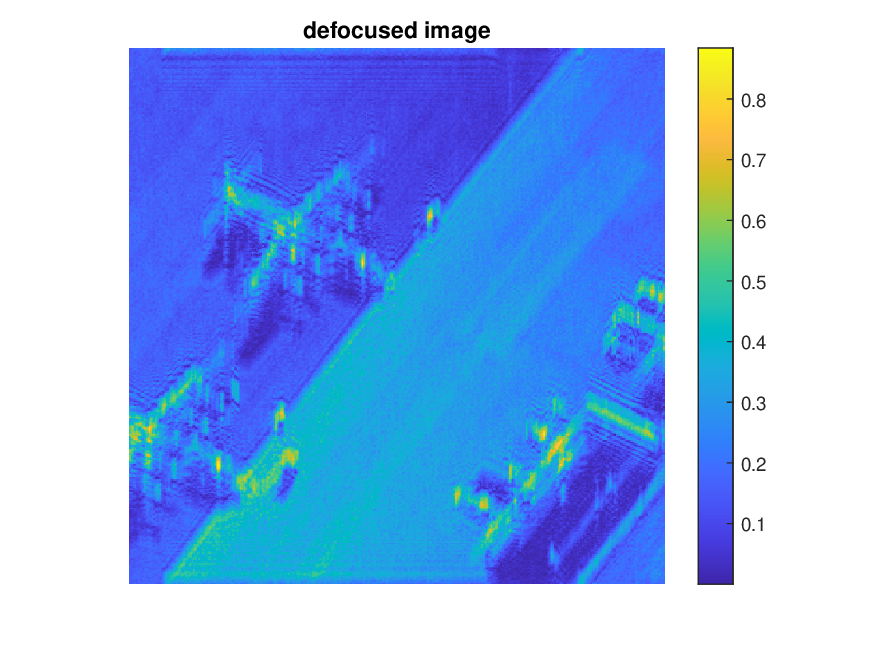}}\\[-2ex] 
	\subfloat[]{\label{fig4c}\includegraphics[width=0.5\linewidth]{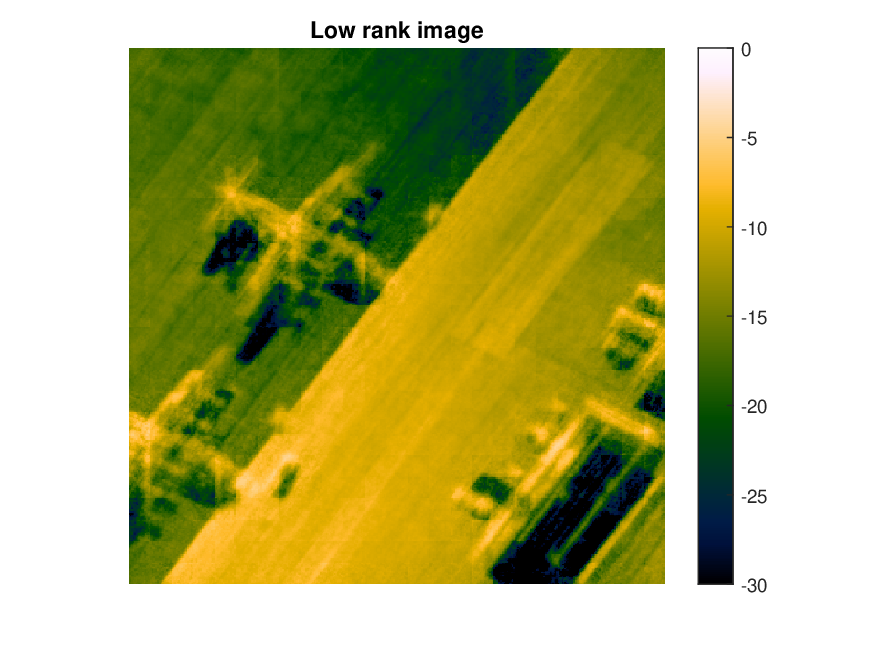}}\hfill
	\subfloat[]{\label{fig4d}\includegraphics[width=0.5\linewidth]{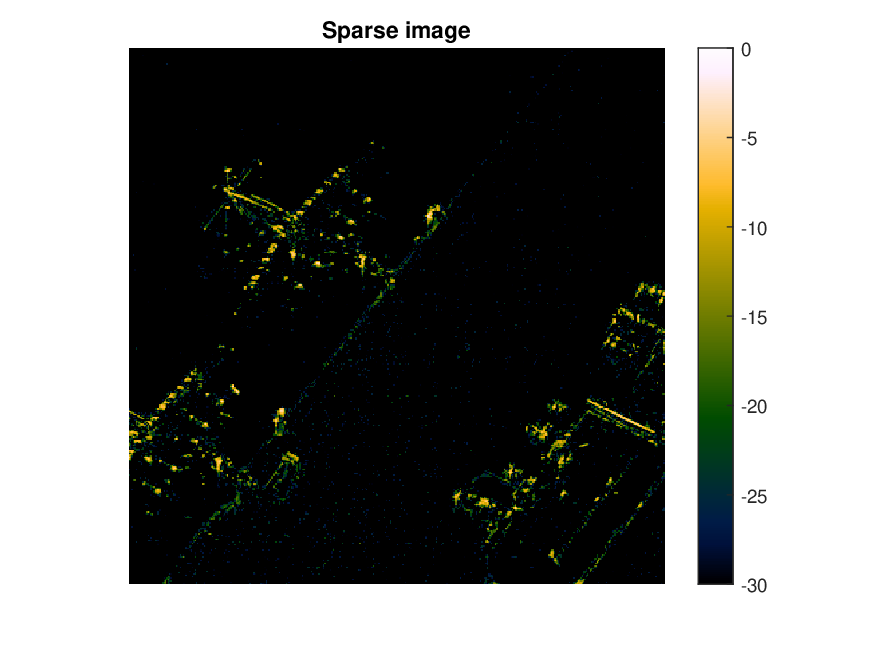}}	 
	\caption{Results of the proposed ADMM-UQP algorithm with airborne ISR dataset
		\protect\subref{fig4a} Original scene. \protect\subref{fig4b} defocused image, \protect\subref{fig4c} low-rank, 
		\protect\subref{fig4d} and sparse images 
	}
	\label{fig4}		       
\end{figure}

In order to show the effectiveness of our proposed framework, we apply our algorithm on a 256 × 256 real SAR scene courtesy of Sandia national laboratories, airborne ISR \cite{Sandia}, and the obtained results are displayed in Fig. \ref{fig4}.  
As in the previous subsection, the proposed method decomposes the low-rank and sparse components successfully.
 
Next, we take advantage of the measured dataset of a Yak-42 aircraft \cite{yak42}, to further	validate the superiority of the proposed approach.
The measured dataset is collected by a radar with a central frequency, bandwidth and PRF of 5.52 GHz, 400 MHz and 100 Hz, respectively. The target is a Yak-42 aircraft of size $24m \times 24m$. The complete radar echo contains
256 pulses, and each pulse consists of 256 samples \cite{Hashempour_sparsity_driv}.

\begin{figure*} 
	
	\medskip
		\begin{minipage}{0.03\textwidth}
			{\small (a)} \\  \\ \\ \\  \\ \\ 
			\\ \\ {\small (b)} \\ \\ \\ \\ \\ 
			\\ \\ \\{\small (c)}
		\end{minipage}\hfill
	\begin{minipage}{0.24\textwidth}
		{ \qquad \qquad \qquad RDA}
		\includegraphics[width=\linewidth]{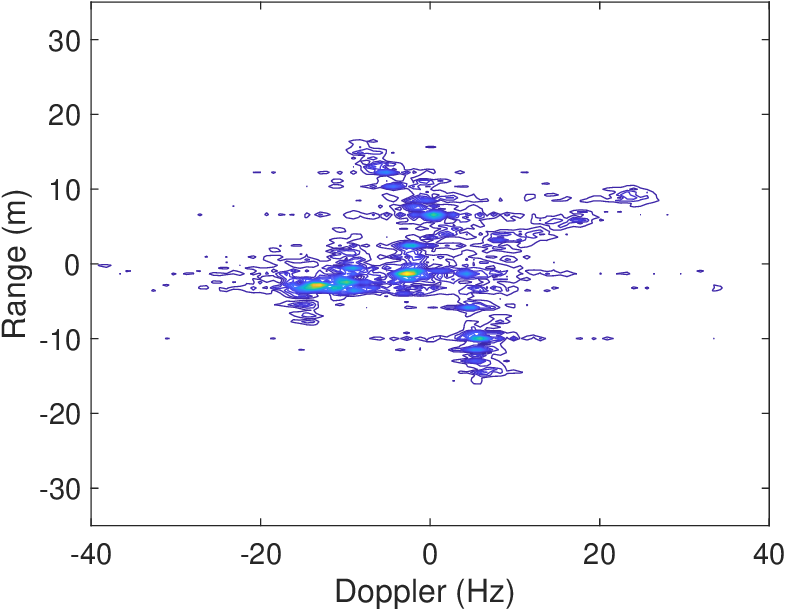}
		\includegraphics[width=\linewidth]{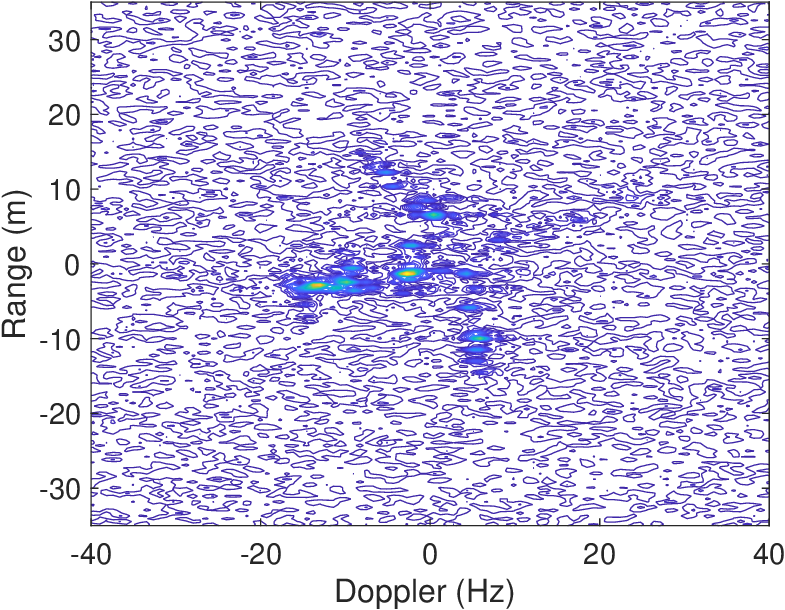}
		\includegraphics[width=\linewidth]{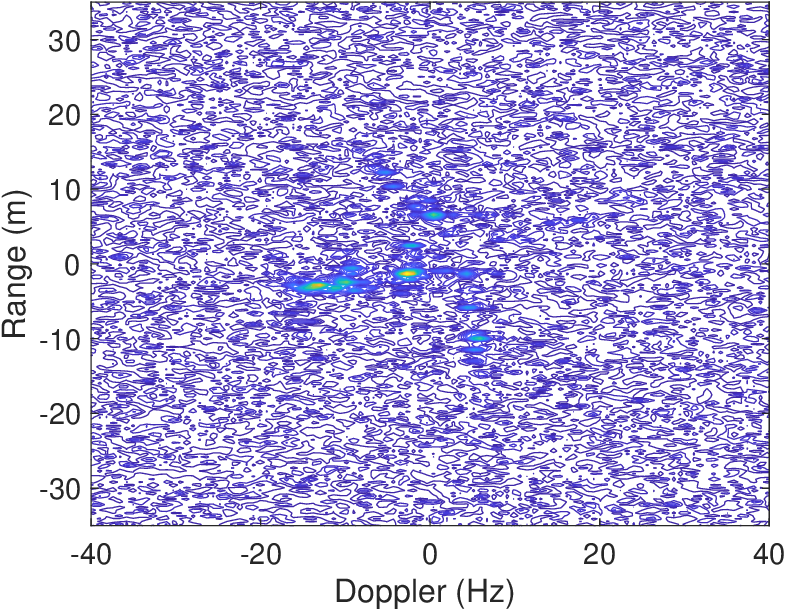}
	\end{minipage}\hfill
	\begin{minipage}{0.24\textwidth}
		{ \qquad \quad \quad ADMM-SBL}
		\includegraphics[width=\linewidth]{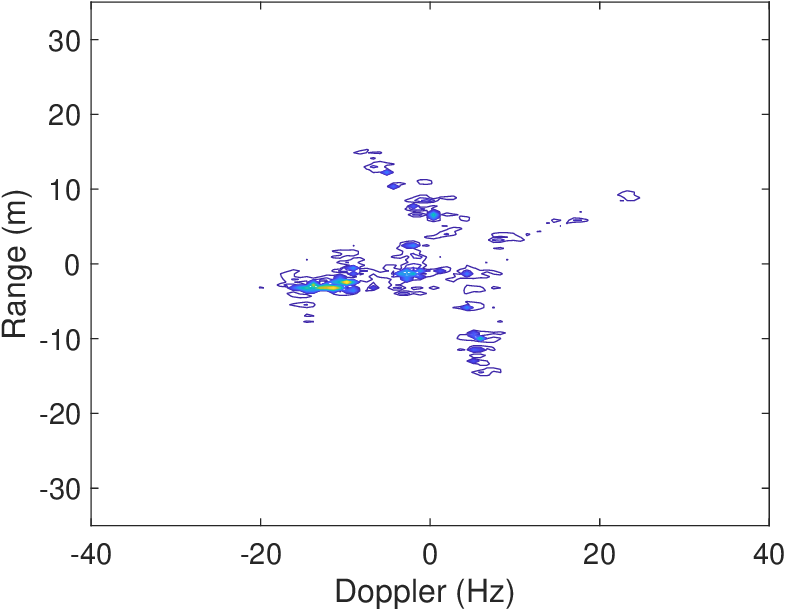}
		\includegraphics[width=\linewidth]{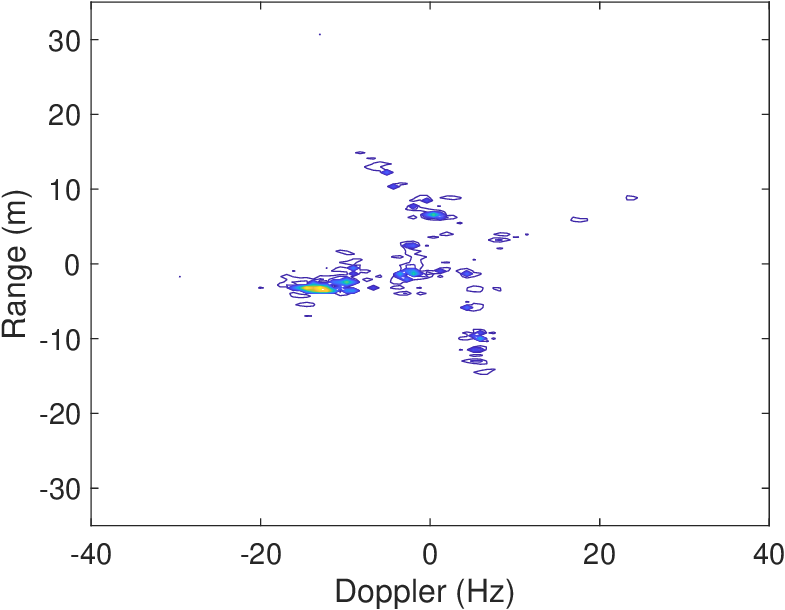}
		\includegraphics[width=\linewidth]{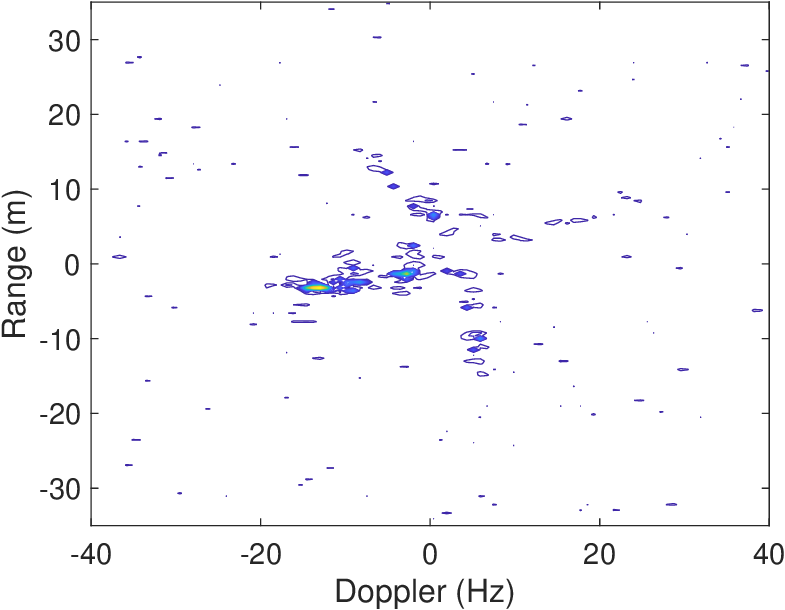}
	\end{minipage}\hfill
	\begin{minipage}{0.24\textwidth}
		{ \qquad \qquad  ADMM-Conv}
			\includegraphics[width=\linewidth]{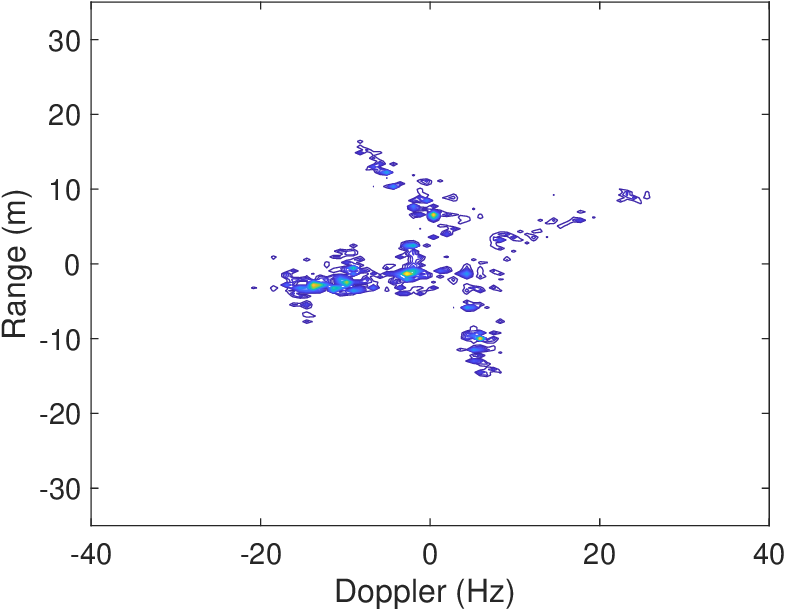}
		\includegraphics[width=\linewidth]{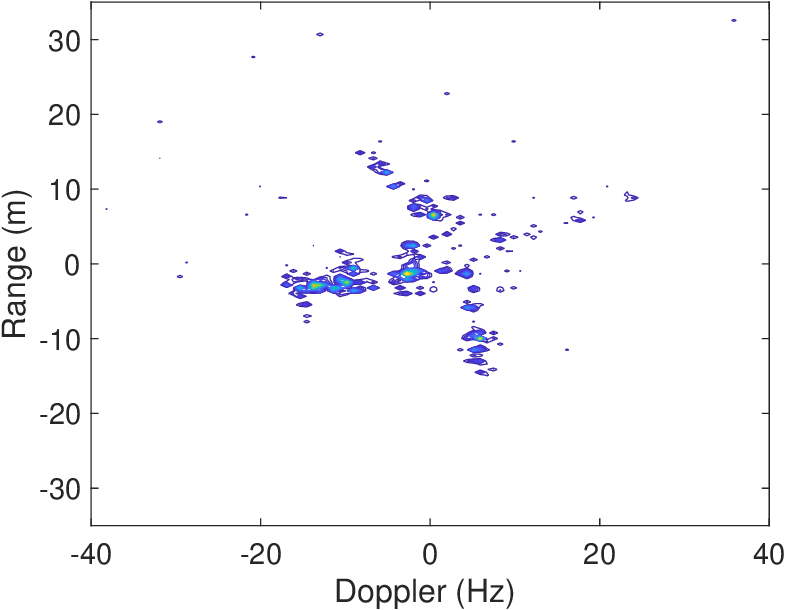}
		\includegraphics[width=\linewidth]{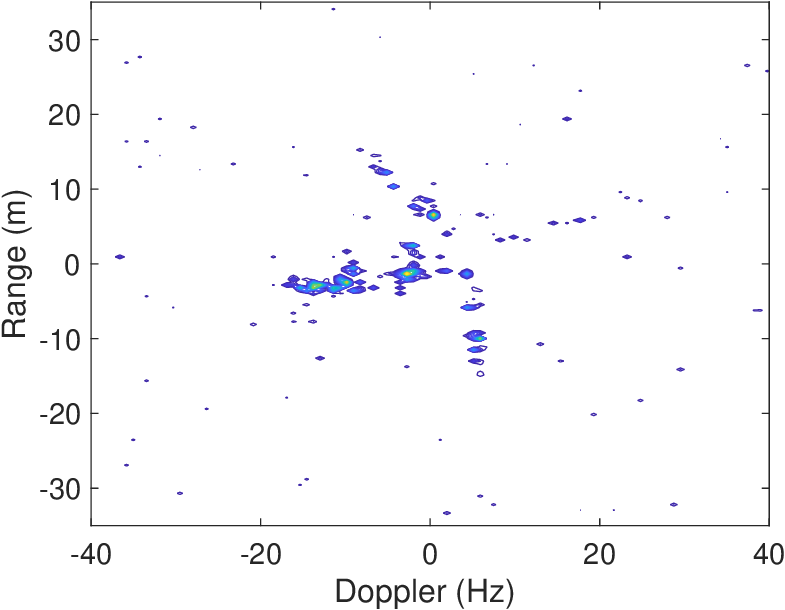}
	\end{minipage}\hfill
	\begin{minipage}{0.24\textwidth}
		{ \qquad \qquad  ADMM-UQP}
		\includegraphics[width=\linewidth]{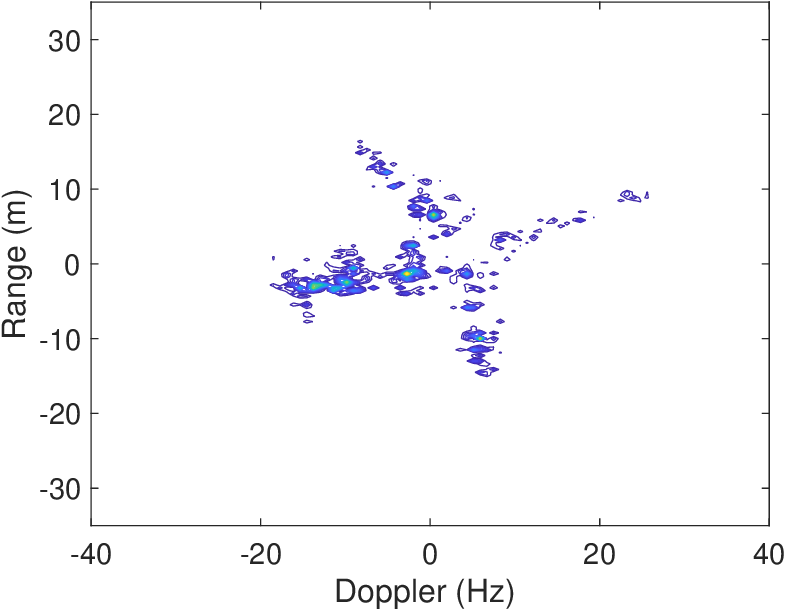}
		\includegraphics[width=\linewidth]{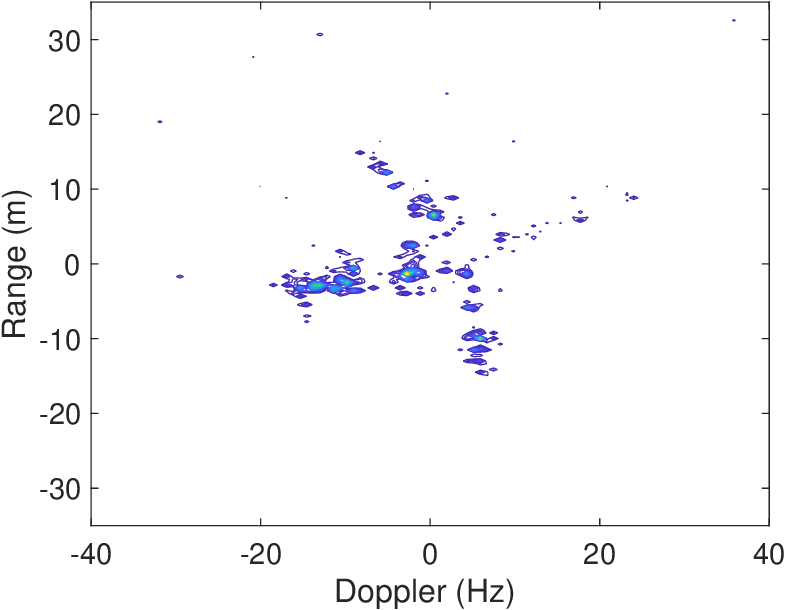}
		\includegraphics[width=\linewidth]{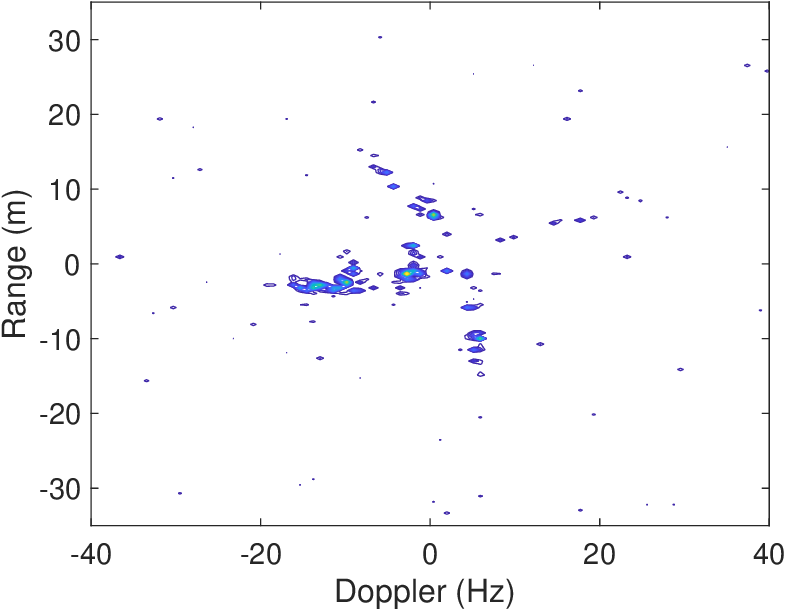}
	\end{minipage}
	\caption{ISAR images of a Yak-42 aircraft \cite{yak42} obtained by different methods for $M_a=64$, $M_r=256$ and with (a) SNR = 10 dB, (b) SNR = 0 dB, and (c) SNR = -5 dB.}
	\label{figg4} 
\end{figure*}
In Fig. \ref{figg4}, the ISAR images of Yak-42 obtained from different methods are demonstrated. The number of pulses $M_a$, and the range cells $M_r$ are 64 and 256, respectively. In order to have super-resolution ISAR images, we set $N_a=2M_a$. The first, second and third rows of Fig. \ref{figg4} are obtained under SNR levels of 10dB, 0dB and -5dB, respectively. From Fig. \ref{figg4}, it can be observed that while the RDA results have the lowest quality, the rest sparsity-driven methods can denoise the images and reconstruct the target even in low SNR conditions. To have a more precise comparison, Table \ref{table5} shows the entropy values of the ISAR images in Fig. \ref{figg4}. Similar to the previous section, the image entropy of ADMM-UQP is the lowest compared to the other algorithms in all SNR regimes.
\begin{table}[!t]
	\small
	\renewcommand{\arraystretch}{1.3}
	\caption{Entropy of the ISAR image of the Yak-42 aircraft \cite{yak42} obtained from different algorithms}
	\centering
	\begin{tabular}{c c c c}
		\hline\hline
		Algorithm & SNR = 10dB &  SNR = 0dB & SNR = -5dB \\
		\hline
		RDA  & 6.12 &  8.29  & 9.32\\
		ADMM-Conv \cite{moradikia} & 4.22   & 3.97 & 3.83\\
		ADMM-SBL \cite{admm-sbl} & 4.35 &   5.03 & 6.70
		\\
		ADMM-UQP &  \textbf{4.11} & \textbf{3.91} & \textbf{3.72}
		\\
		\hline		
	\end{tabular}
	\label{table5}
\end{table}

Table \ref{run_time2} compares the run-time of different methods. As expected, the proposed ADMM-UQP method converges faster than ADMM-Conv and ADMM-SBL in all SNR settings. Therefore, both entropy and run-time results confirm the superiority of the proposed algorithm.  
\begin{table}[!t]
	\small
	\renewcommand{\arraystretch}{1.3}
	\caption{Run-time of different algorithms for real data}
	\centering
	\begin{tabular}{c c c c}
		\hline\hline
		Algorithm & SNR = 10dB &  SNR = 0dB & SNR = -5dB \\
		\hline
		ADMM-Conv \cite{moradikia} & 14.01s   & 14.30s & 14.10s\\
		ADMM-SBL \cite{admm-sbl} & 19.10s &   18.80s & 16.88s
		\\
		ADMM-UQP &  \textbf{6.50s} & \textbf{6.45s} & \textbf{6.48s}
		\\
		\hline
		
	\end{tabular}
	\label{run_time2}
\end{table}

\begin{figure*} 
	
	\medskip
	\begin{minipage}{0.03\textwidth}
		{\small (a)} \\  \\ \\ \\  \\ \\ 
		\\ \\ {\small (b)} \\ \\ \\ \\ \\ 
		\\ \\ \\{\small (c)}
	\end{minipage}\hfill
	\begin{minipage}{0.24\textwidth}
		{ \qquad \qquad \qquad HRRP}
		\includegraphics[width=\linewidth]{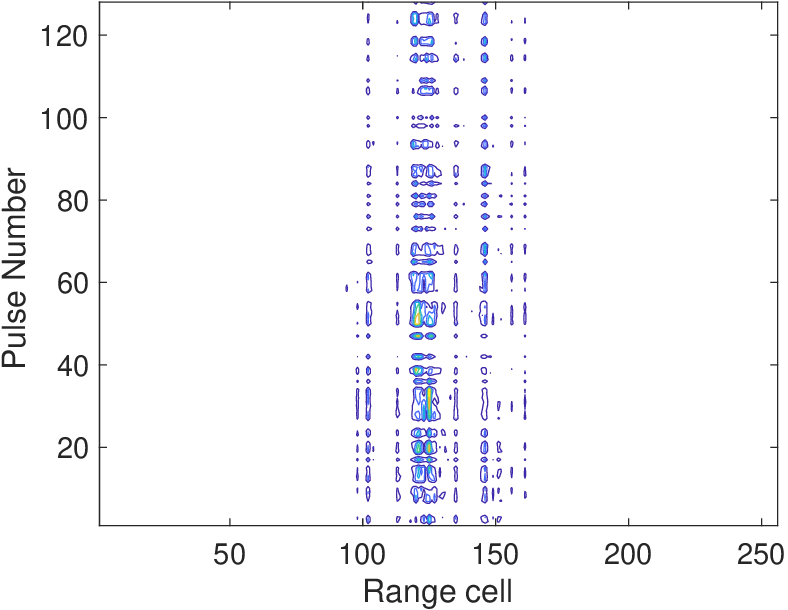}
		\includegraphics[width=\linewidth]{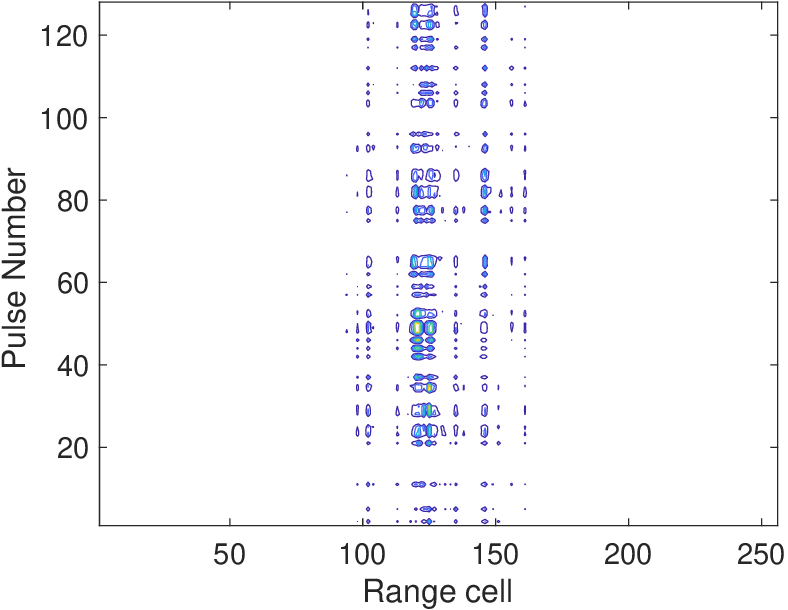}
		\includegraphics[width=\linewidth]{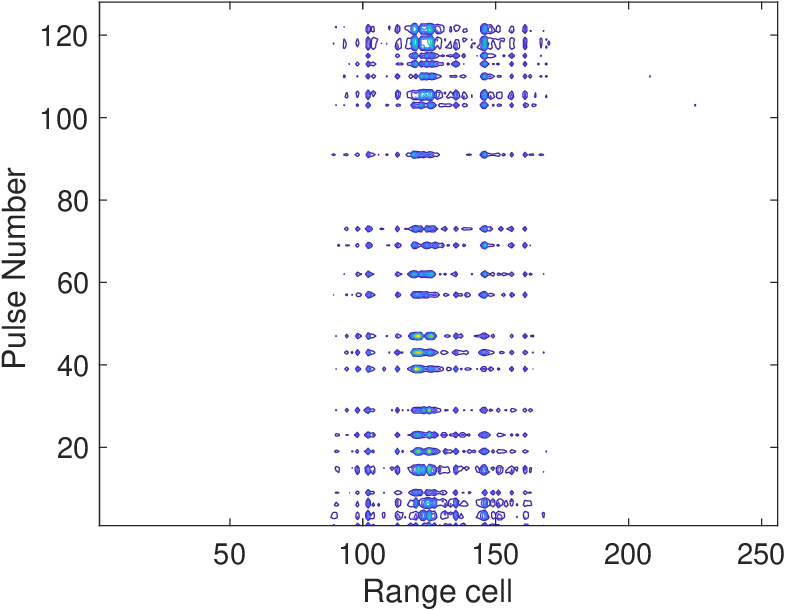}
	\end{minipage}\hfill
	\begin{minipage}{0.24\textwidth}
		{ \qquad \quad \quad ADMM-SBL}
		\includegraphics[width=\linewidth]{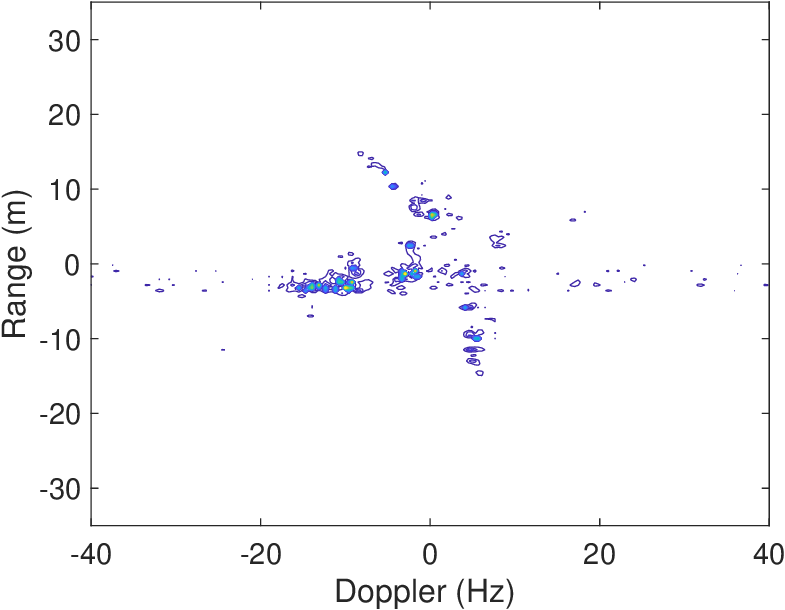}
		\includegraphics[width=\linewidth]{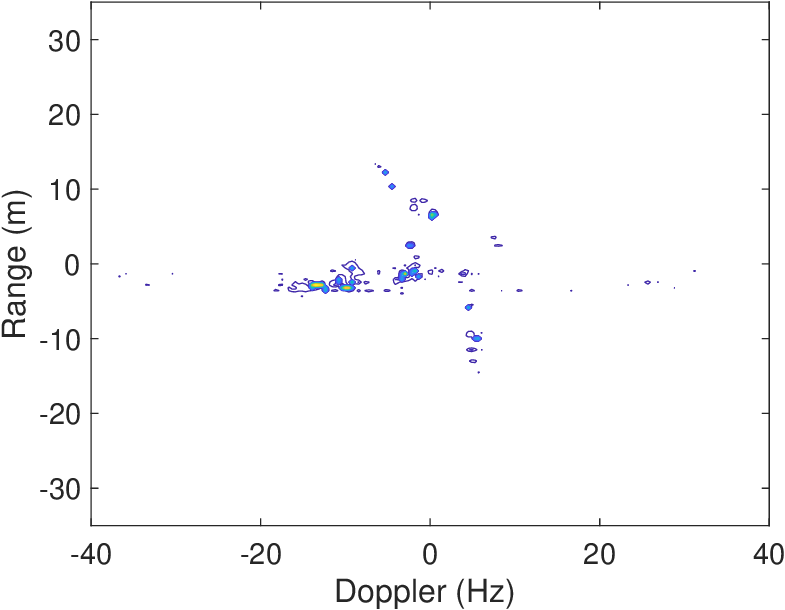}
		\includegraphics[width=\linewidth]{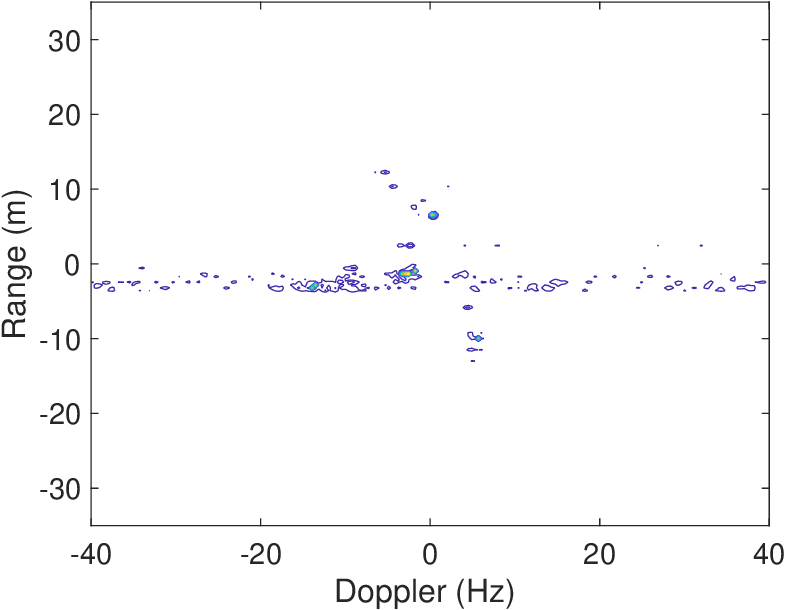}
	\end{minipage}\hfill
	\begin{minipage}{0.24\textwidth}
		{ \qquad \quad \quad ADMM-Conv}
		\includegraphics[width=\linewidth]{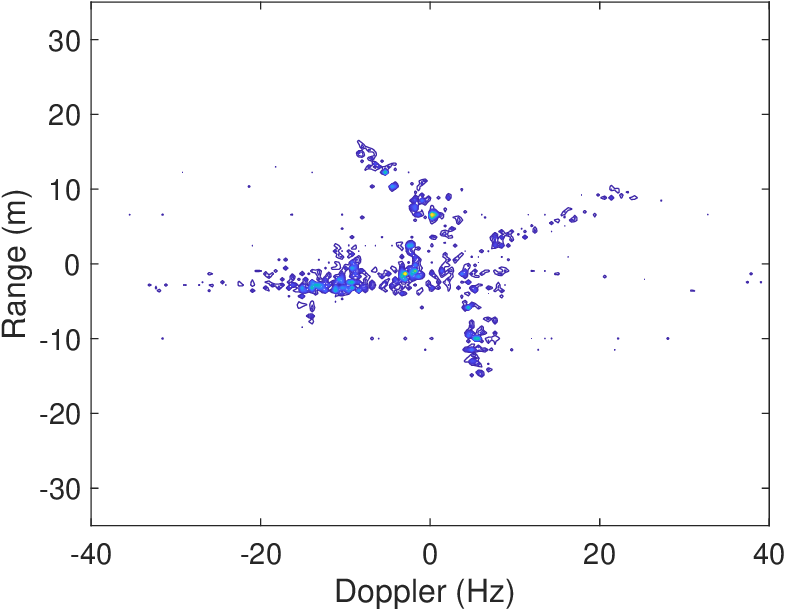}
		\includegraphics[width=\linewidth]{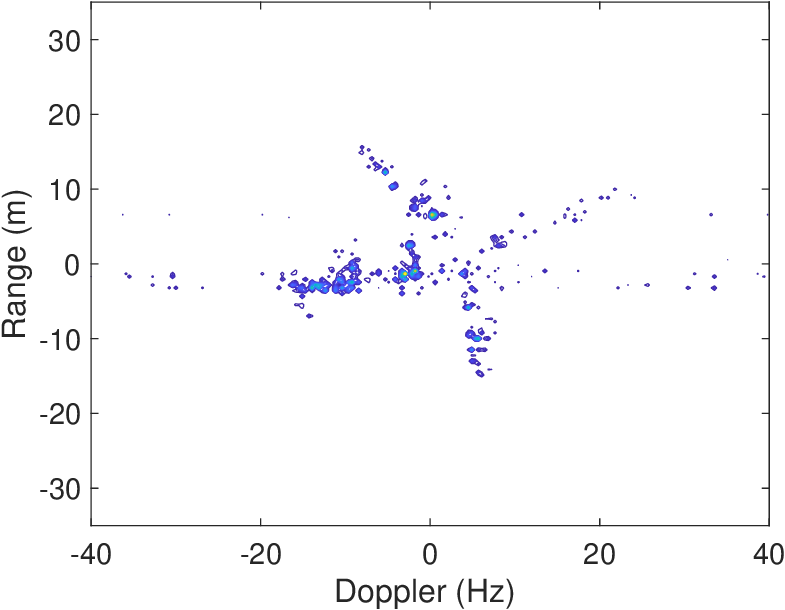}
		\includegraphics[width=\linewidth]{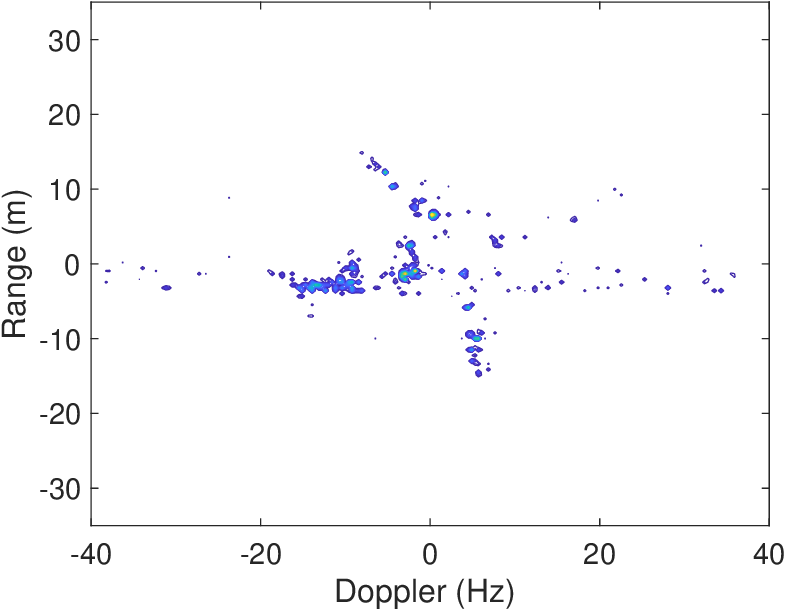}
	\end{minipage}\hfill
	\begin{minipage}{0.24\textwidth}
		{ \qquad \qquad  ADMM-UQP}
		\includegraphics[width=\linewidth]{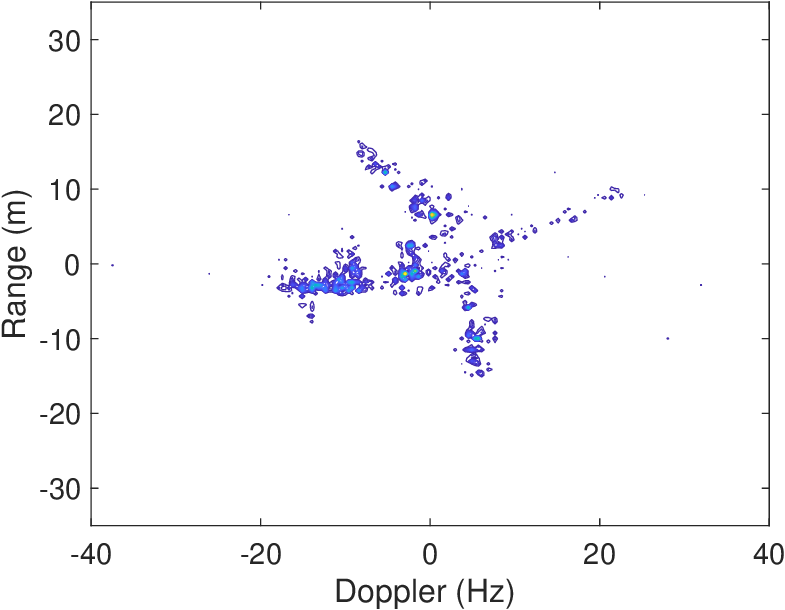}
		\includegraphics[width=\linewidth]{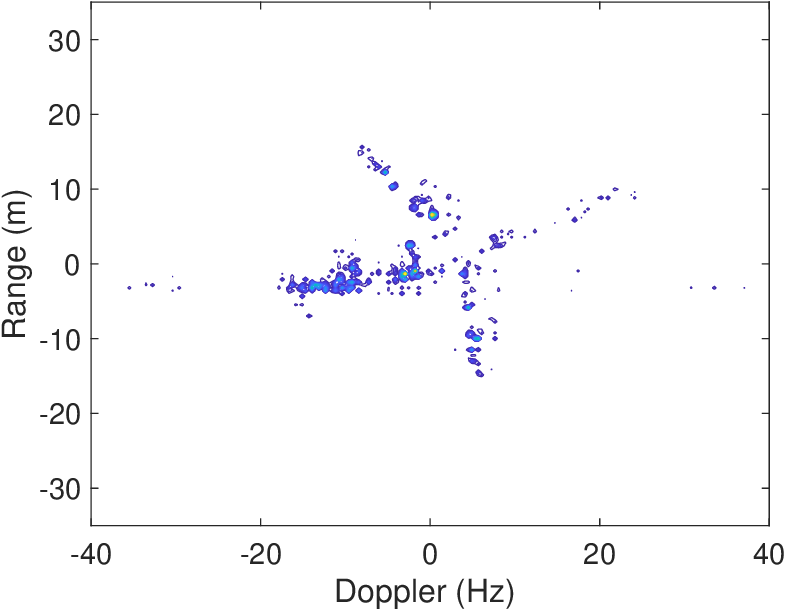}
		\includegraphics[width=\linewidth]{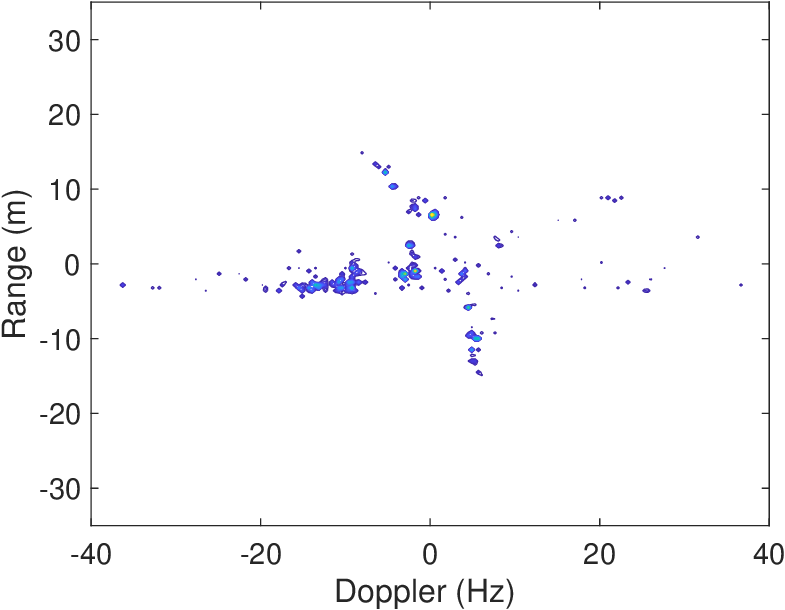}
	\end{minipage}
	\caption{Range profile and ISAR images of a Yak-42 aircraft \cite{yak42} obtained by different methods with the udersampling ratio of (a) 0.75, (b) 0.50, and (c) 0.25.}
	\label{figg5} 
\end{figure*}

For our final experiment, we evaluate the performance of the proposed algorithm with the undersampled data. Assume that the full pulse number is 128, and we select 96, 64 and 32 pulses in the high resolution range profile (HRRP) randomly, equivalent to 0.75, 0.5 and 0.25 undersampling schemes, respectively. Fig. \ref{figg5} shows the HRRP and the achieved results by different algorithms. Artificially induced points emerge in the images obtained from all methods specially for lower undersampling ratios. However, ADMM-UQP appears to achieve better results compared to other approaches.

\section{Conclusion}\label{conc}
In this paper, an ADMM-UQP method for robust LRSD-based SAR/ISAR imaging was proposed, and was shown to be more computationally efficient than state-of-the-art methods and to require lower memory usage. A novel UQP-based approach is used for autofocusing that achieves faster convergence than the conventional methods.
Moreover, the performance of the proposed approach appears to be better than RDA, ADMM-SBL and ADMM-Conv methods, at different
SNR levels and various undersampling conditions which further validates the robustness of our algorithm. Both synthetic and
real-world data were used to show the superiority of the proposed method in terms of quality and computational cost. 


%

\appendices
\section{Proof of the Relations in Section \ref{sec3a}}
Let ${\mathbf{F}}_a \in {\mathbb{C}}^{N_a\times \sqrt{I}}$ and ${\mathbf{F}}_r\in{\mathbb{C}}^{N_r\times \sqrt{I}}$ denote the corresponding partial 1-D Fourier transform matrix in azimuth and range directions with $N_a\mathrm{\ and}\ N_r\ll \sqrt{I}$ to create a super-resolution image, respectively. As the observation kernel $\mathbf{H}$ can be approximated by the 2D-Fourier matrix, we can re-express it through the Kronecker product decomposition (KPD) as follows:
\begin{align}
\mathbf{H}={\mathbf{F}}_r \otimes {\mathbf{F}}_a,
\end{align}
where the definitions of ${\mathbf{F}}_a$ and ${\mathbf{F}}_r$ are given as:

\begin{align}
\mathbf{F}_a \triangleq &
\begin{bmatrix}
1      & 1     & \dots  & 1 \\ 
1     & \omega   & \dots & \omega^{(I_a-1)}\\
\vdots  &\vdots  & \ddots &\vdots\\	
1 & \omega^{(N_a-1)}& \dots  & \omega^{(N_a-1)(I_a-1)}	     \\
\end{bmatrix},
\end{align}
\begin{align}
\mathbf{F}_r \triangleq
\begin{bmatrix}
1      & 1     & \dots  & 1 \\ 
1     & \nu   & \dots & \nu^{(I_r-1)}\\
\vdots  &\vdots  & \ddots &\vdots\\	
1 & \nu^{(N_r-1)}& \dots  & \nu^{(N_r-1)(I_r-1)}	     \\
\end{bmatrix}, 
\end{align}
where $ \omega \triangleq \exp {(-j\frac{2\pi}{N_a})} $ and $ \nu \triangleq \exp {(-j\frac{2\pi}{M})} $. Interestingly, as ${\mathbf{F}}_a{\mathbf{F}}^H_a={\mathbf{F}}_r{\mathbf{F}}^H_r=\mathbf{I}$ we can simply prove that $\boldsymbol{\mathrm{H}}{\boldsymbol{\mathrm{H}}}^H=\boldsymbol{\mathrm{I}}$, since:

\begin{align}
\mathbf{H}{\mathbf{H}}^H &= \left({\mathbf{F}}_r \otimes {\mathbf{F}}_a\right){\left({\mathbf{F}}_r \otimes {\mathbf{F}}_a\right)}^H \nonumber \\ &=\left({\boldsymbol{\mathrm{F}}}_r\otimes {\boldsymbol{\mathrm{F}}}_a\right)\left({\boldsymbol{\mathrm{F}}}^H_r \otimes {\boldsymbol{\mathrm{F}}}^H_a\right)
\nonumber \\ &
=\left({\mathbf{F}}_r {\mathbf{F}}^H_r \right)\otimes \left({\mathbf{F}}_a {\mathbf{F}}^H_a\right)
\nonumber \\ &=\mathbf{I}.
\end{align}
On the other hand, since the under-sampling matrix $\boldsymbol{\mathrm{\Theta }}$ is obtained by selecting a subset of rows of an identity matrix, we have $\boldsymbol{\mathrm{\Theta }}{\boldsymbol{\mathrm{\Theta }}}^T = \mathbf{I}$ and thus $\mathbb{E}{\mathbb{E}}^H=\mathbf{I}$. Now, using the matrix inversion lemma \cite{Hashempour_sparsity_driv} the terms (a), (b) in \eqref{EQ8}, \eqref{EQ9} can be equivalently written as
\begin{align}\label{EQ26}
{\left({\mathbb{E}}^{H}\mathbb{E}+{\delta }_{i}\mathbf{I}\right)}^{-1}=\frac{1}{{\delta }_{i}}\left(\mathbf{I}-\frac{1}{1+{\delta }_{i}}{\mathbb{E}}^H\mathbb{E}\right), \ i \in \left\{\mathrm{1,2}\right\}.
\end{align}
Substituting \eqref{EQ26} in \eqref{EQ8} and \eqref{EQ9} and we obtain \eqref{EQ13} and \eqref{EQ14}, respectively. The proof of \eqref{EQ15} is also straightforward and can be reached through trivial algebraic manipulations. 

\section{Derivations of Relations \eqref{EQ17} and \eqref{EQ18} in Section \ref{sec3b}}
 Given the matrix model in \eqref{EQ16}, the term (d)  in \eqref{EQ13} defined as $\left(d\right)\triangleq {\mathbb{E}}^H\mathbb{E}\left\{ \mathbf{\Lambda} \right\}-{\mathbb{E}}^H\mathbf{r}$ with  $ \boldsymbol{\Lambda}\triangleq \mathbf{W}-{\delta }^{-1}_1{\mathbf{Z}}_{1}+\mathbf{S}$
  can be recast as \eqref{EQ27} at the top of next page.

\begin{figure*}
	\begin{align}\label{EQ27}
	(d)&=\mathcal{P}\left\{\left[{\left({\boldsymbol{\Theta }}_{r}{\mathbf{F}}_{r}\right)}^H\otimes \ {\left({\boldsymbol{\Phi }\boldsymbol{\Theta }}_{a}{\mathbf{F}}_{a}\right)}^H\right]\left[\left({\boldsymbol{\Theta }}_{r}{\mathbf{F}}_{r}\right)\otimes \left({\boldsymbol{\Phi }\boldsymbol{\Theta }}_{a}{\mathbf{F}}_{a}\right)\right]{\mathcal{P}}^{-1}\left(\mathrm{vec}\left\{\boldsymbol{\Lambda }\right\}\right)-{\left({\boldsymbol{\Theta }}_{r}{\mathbf{F}}_{r}\right)}^H\otimes \ {\left({\boldsymbol{\Theta }}_{a}{\mathbf{F}}_{a}\right)}^H{\boldsymbol{\Phi }}^{H}\mathrm{vec}\{\mathbf{R}\}\right\}
	\nonumber \\
	&=\mathcal{P}\left\{\left[\left({\mathbf{F}}_{r}^H{\boldsymbol{\Theta }}^T_{r}{\boldsymbol{\Theta }}_{r}\mathbf{F}_{r}\right)\otimes \left({\mathbf{F}}^H_{a}{\boldsymbol{\Theta }}^T_{a}{\boldsymbol{\Phi }}^{H}\boldsymbol{\Phi }{\boldsymbol{\Theta }}_{a}{\boldsymbol{F}}_{a}\right)\right]{\mathcal{P}}^{-1}\left(\mathrm{vec}\left\{\boldsymbol{\Lambda }\right\}\right)-{\mathbf{F}}_{r}^H{\boldsymbol{\Theta }}^T_{r}\otimes \ {\boldsymbol{\mathrm{F}}}^H_{a}{\boldsymbol{\Theta }}^T_{a}{\boldsymbol{\Phi }}^{H}\boldsymbol{ }\mathrm{vec}\left\{\boldsymbol{\mathrm{R}}\right\}\right\}
	\nonumber	\\ 
	&
	=\mathcal{G}\left\{{\boldsymbol{\mathrm{F}}}^H_{a}{\boldsymbol{\mathrm{\Theta }}}^T_{a}{\boldsymbol{\mathrm{\Phi }}}^{H}\boldsymbol{\mathrm{\Phi }}{\boldsymbol{\mathrm{\Theta }}}_{a}{\boldsymbol{\mathrm{F}}}_{a}{\mathcal{G}}^{-1}\left\{\boldsymbol{\mathrm{\Lambda }}\right\}\mathbf{F}_{r}^T{\boldsymbol{\mathrm{\Theta }}}^T_{r}{\boldsymbol{\mathrm{\Theta }}}_{r}{\boldsymbol{\mathrm{F}}}^{*}_{r}-{\boldsymbol{\mathrm{F}}}^H_{a}{\boldsymbol{\mathrm{\Theta }}}^T_{a}{\boldsymbol{\Phi}}^{H}\mathbf{R}{\boldsymbol{\Theta }}_{r}{\mathbf{F}}^{*}_{r}\right\}
	\nonumber \\
	&=\mathcal{G}\left\{{\mathbf{F}}^H_{a}{\boldsymbol{\Theta }}^T_{a}{\boldsymbol{\Phi }}^H\left[\boldsymbol{\Phi }{\boldsymbol{\Theta }}_{a}{\mathbf{F}}_{a}{\mathcal{G}}^{-1}\left\{\boldsymbol{\Lambda }\right\}{\mathbf{F}}_{r}^T{\boldsymbol{\Theta }}^T_{r}-\mathbf{R}\right]{\boldsymbol{\Theta }}_{r}\mathbf{F}^{*}_{r}\right\}.
	\end{align}
	\hrulefill
\end{figure*}
By substituting the above matrix representation of the term (d) into \eqref{EQ13}, we obtain \eqref{EQ17}. We can follow a similar approach to derive \eqref{EQ18}. 

\section{Effectveness of the UQP formulation and power method-like iterations}
We show herein that the iterations in \eqref{EQ23} leads to a monotonic increase of the UQP objective function \eqref{EQ22}. To do so, we first note that the problem \eqref{EQ22}, after some mathematical manipulations, can be transformed into

\begin{align}\label{EQ28}
\widehat{\mathbf{p}}={\mathrm{arg} {\mathop{\mathrm{max}}_{\mathbf{p}}   \ {\widetilde{\mathbf{p}}}^{{\left(t+1\right)}^{H}}\mathbf{U}{\widetilde{\mathbf{p}}}^{(t)}\ }\ }, \ \  \mathrm{s.t.}\ \ \left|{\left(\widetilde{\mathbf{p}}\right)}_i\right|=1, \ \ \forall i.
\end{align}
At the $\left(t+1\right)$th iteration, if we assume ${\widetilde{\mathbf{p}}}^{\left(t\right)}$ is fixed, the updated ${\widetilde{\mathbf{p}}}^{\left(t+1\right)}$ acquired by \eqref{EQ23} is the solution to the maximization problem \eqref{EQ28}. Therefore, since $\mathbf{U}$ is positive definite we have
\begin{align}
{\left({\widetilde{\mathbf{p}}}^{\left(t+1\right)}-{\widetilde{\mathbf{p}}}^{(t)}\right)}^H\mathbf{U}\left({\widetilde{\mathbf{p}}}^{\left(t+1\right)}-{\widetilde{\mathbf{p}}}^{(t)}\right) \ge 0,
\end{align}
which implies
\begin{align}
{\widetilde{\mathbf{p}}}^{{\left(t+1\right)}^{H}}&\mathbf{U}{\widetilde{\mathbf{p}}}^{(t+1)}
\nonumber\\
&\ge 2\mathfrak{R}\left\{{\widetilde{\mathbf{p}}}^{{\left(t+1\right)}^{H}}\mathbf{U}{\widetilde{\mathbf{p}}}^{\left(t\right)}\right\}-
{\widetilde{\mathbf{p}}}^{{\left(t\right)}^{H}}\mathbf{U}{\widetilde{\mathbf{p}}}^{\left(t\right)}
\nonumber\\
&\overbrace{\ge }^{\left(*\right)}
{\widetilde{\mathbf{p}}}^{{\left(t\right)}^{H}}\mathbf{U}{\widetilde{\mathbf{p}}}^{\left(t\right)}.
\end{align}
Note that (*) holds due to the fact that $\mathfrak{R}\left\{{\widetilde{\mathbf{p}}}^{{\left(t+1\right)}^{H}}\mathbf{U}{\widetilde{\mathbf{p}}}^{\left(t\right)}\right\}\ge {\widetilde{\mathbf{p}}}^{{\left(t\right)}^{H}}\mathbf{U}{\widetilde{\mathbf{p}}}^{(t)}$.

\begin{IEEEbiography}[{\includegraphics[width=1in,height=1.25in,clip,keepaspectratio]{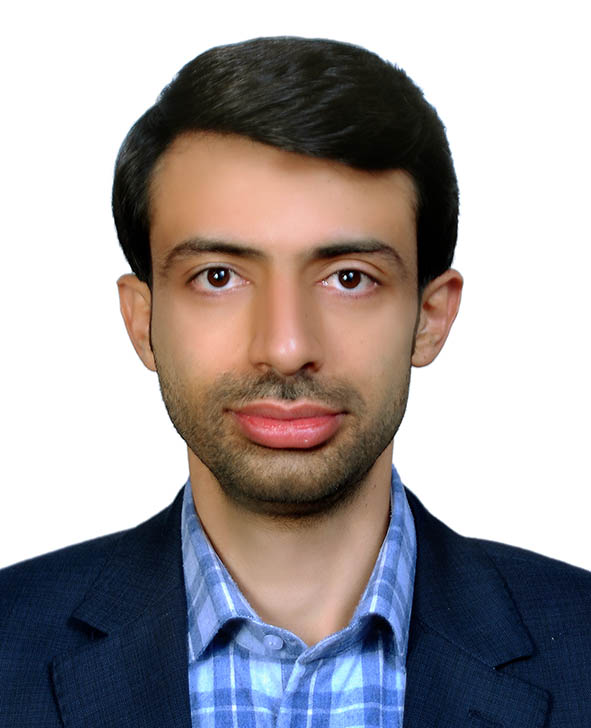}}]{Hamid Reza Hashempour}
	was born in 1987. He received the B.S., M.S., and Ph.D. degrees in electrical engineering from Shiraz University, Shiraz, Iran in 2009, 2011, and 2017, respectively.
	He is currently a Research Assistant with the Electrical Engineering Department, Sharif University of Technology, Tehran, Iran. 
	His current research interests include radar signal
	processing, radar imaging (SAR/ISAR), compressive sensing and physical-layer security of wireless communications.
\end{IEEEbiography}
\begin{IEEEbiography}[{\includegraphics[width=1in,height=1.25in,clip,keepaspectratio]{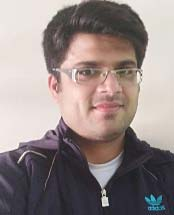}}]{Majid Moradikia}
	was born in 1986. He received the
	Ph.D. degree in telecommunication system engineering from the Department of Electrical and Electronics Engineering, Shiraz University of Technology,
	Shiraz, Iran. He currently works as a Postdoctoral Research Fellow with the Engineering Technology Department, University of Houston, TX, USA.
	His main research interests lie within the area of
	physical-layer security of wireless communications,
	the Internet of Things, millimeter-wave communication systems, and massive MIMO systems.
\end{IEEEbiography}
\begin{IEEEbiography}[{\includegraphics[width=1in,height=1.25in,clip,keepaspectratio]{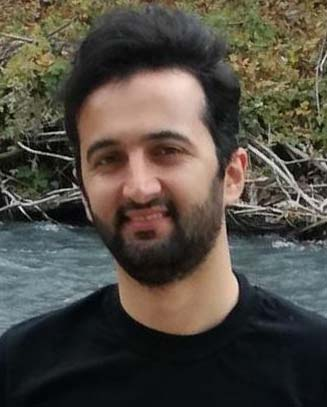}}]{Hamed Bastami}
	received the M.Sc. degree in electrical engineering from the University of Tehran,
	Tehran, Iran, in 2014. He is currently pursuing
	the Ph.D. degree with the Department of Electrical
	Engineering, Sharif University of Technology. His
	research interests lie in the areas of physical-layer
	security (PLS) of wireless communications with
	special emphasis on machine learning, deep-based
	resource management, PLS in wireless communication, and fog radio access networks.
\end{IEEEbiography}
\begin{IEEEbiography}[{\includegraphics[width=1in,height=1.25in,clip,keepaspectratio]{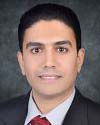}}]{Ahmed Abdelhadi}
	(Senior Member, IEEE) received
	the Ph.D. degree in electrical and computer engineering from The University of Texas at Austin
	in 2011. He is currently an Assistant Professor
	with the University of Houston. Before joining UH,
	he was a Research Assistant Professor with Virginia
	Tech. He was a member of the Wireless Networking and Communications Group (WNCG) and the
	Laboratory of Informatics, Networks and Communications (LINC) Group during his Ph.D. In 2012,
	he joined the Bradley Department of Electrical and
	Computer Engineering and the Hume Center for National Security and
	Technology, Virginia Tech. He was a Faculty Member of Wireless @ Virginia
	Tech. His research interests are in the areas of wireless communications
	and networks, artificial intelligence, cyber physical systems, and security.
	He has coauthored more than 80 journal and conference papers, and seven
	books in these research topics. His book Cellular Communications Systems
	in Congested Environments is bookplated as the Virginia Tech Provost’s
	Honor Book and his book Resource Allocation with Carrier Aggregation
	in Cellular Networks is featured in the 13th Annual Virginia Tech Authors
	Recognition Event. He received the Silver Contribution Award from the IEEE
	International Conference on Computing, Networking and Communications
	(ICNC), the Best Paper Award from the IEEE International Symposium on
	Systems Engineering (ISSE), and the Outstanding Paper Award from the IEEE
	International Conference of Advanced Communications Technology (ICACT).
\end{IEEEbiography}
\begin{IEEEbiography}[{\includegraphics[width=1in,height=1.25in,clip,keepaspectratio]{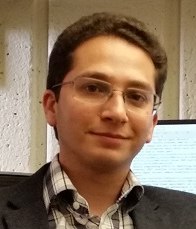}}]{Mojtaba Soltanalian}
 (Senior Member, IEEE) received the Ph.D. degree in electrical engineering (with specialization in signal processing) from the Department of Information Technology, Uppsala University, Sweden, in 2014. He is currently with the faculty of the Electrical and Computer Engineering Department, University of Illinois Chicago (UIC), Chicago, IL, USA. Before joining UIC, he held research positions with the Interdisciplinary Centre for Security, Reliability and Trust (SnT, University of Luxembourg), and California Institute of Technology, Pasadena, CA, USA. His research interests include interplay of signal processing, learning and optimization theory, and specifically different ways the optimization theory can facilitate a better processing and design of signals for collecting information, communication, as well as to form a more profound understanding of data, whether it is in everyday applications or in large-scale, complex scenarios. Dr. Soltanalian serves as an Associate Editor for IEEE Transactions on Signal Processing and as the Chair of the IEEE Signal Processing Society Chapter in Chicago. He was the recipient of the 2017 IEEE Signal Processing Society Young Author Best Paper Award, and the 2018 European Signal Processing Association Best PhD Award.
\end{IEEEbiography}

\end{document}